\documentclass[useAMS,usenatbib]{mn2e}

\usepackage{graphicx}
\usepackage{amssymb}
\usepackage{amsmath}
\usepackage{datetime}
\usepackage[normalem]{ulem}
\usepackage{times}
\usepackage{hyperref}

\usepackage{color}
\usepackage{soul}
\definecolor{stan}{rgb}{1,0,0}

\date{Accepted 2015 August 10.  Received 2015 August 10; in original form 2015 June 3}

\begin{document}

\title[Solution Topology of Line-Driven Winds]
{
Effect of scattering on the transonic solution topology and intrinsic variability of line-driven stellar winds
}

 \author[Sundqvist \& Owocki]
{
\vbox{
Jon O. Sundqvist$^1$\thanks{Email: mail@jonsundqvist.com} \&
Stanley P. Owocki$^1$
}
\\ $^1$ Department of Physics and Astronomy, Bartol Research Institute,
 University of Delaware, Newark, DE 19716, USA
}

\def\<<{{\ll}}
\def\>>{{\gg}}
\def\wig{{\sim}}
\def\spose#1{\hbox to 0pt{#1\hss}}
\def\ltwig{\mathrel{\spose{\lower 3pt\hbox{$\mathchar"218$}}
     R_{\rm A}ise 2.0pt\hbox{$\mathchar"13C$}}}
\def\gtwig{\mathrel{\spose{\lower 3pt\hbox{$\mathchar"218$}}
     R_{\rm A}ise 2.0pt\hbox{$\mathchar"13E$}}}
\def\+/-{{\pm}}
\def\=={{\equiv}}
\def\mubar{{\bar \mu}}
\def\mustar{\mu_{\ast}}
\def\Lambar{{\bar \Lambda}}
\def\Rstar{R_{\ast}}
\def\Bstar{B_{\ast}}
\def\Mstar{M_{\ast}}
\def\Lstar{L_{\ast}}
\def\Tstar{T_{\ast}}
\def\gstar{g_{\ast}}
\def\vth{v_{\rm th}}
\def\grad{g_{\rm rad}}
\def\glines{g_{lines}}
\def\Mdot{\dot M}
\def\mdot{\dot m}
\def\yr{{\rm yr}}
\def\ksec{{\rm ksec}}
\def\kms{{\rm km/s}}
\def\qad{\dot q_{ad}}
\def\qlines{\dot q_{lines}}
\def\solar{\odot}
\def\Msun{M_{\solar}}
\def\msbyr{\Msun/\yr}
\def\Rsun{R_{\solar}}
\def\Lsun{L_{\solar}}
\def\Be{{\rm Be}}
\def\Rpole{R_{p}}
\def\Req{R_{eq}}
\def\Rmin{R_{min}}
\def\Rmax{R_{max}}
\def\Rstag{R_{stag}}
\def\vinf{V_\infty}
\def\Vrot{V_{rot}}
\def\Vcrit{V_{\rm crit}}
\def\half{{\frac{1}{2}}}
\newcommand{\beq}{\begin{equation}}
\newcommand{\eeq}{\end{equation}}
\newcommand{\beqa}{\begin{eqnarray}}
\newcommand{\eeqa}{\end{eqnarray}}
\def\phip{{\phi'}}
\newcommand{\erf}{\mathrm{erf}\,}
\newcommand{\erfc}{\mathrm{erfc}\,}

\maketitle

\begin{abstract}

For line-driven winds from hot, luminous OB stars, we examine the subtle but
important role of diffuse, scattered radiation in determining both the topology of
steady-state solutions and intrinsic variability in the transonic wind
base. We use a smooth source function formalism to obtain nonlocal, 
integral expressions for the direct and diffuse components of the line-force
that account for deviations from the usual localized, Sobolev
forms. As the scattering source function is reduced, we find the
solution topology in the transonic region transitions from X-type,
with a unique wind solution, to a \textit{nodal} type, characterized by a
\textit{degenerate family} of solutions.

Specifically, in the idealized case of an optically thin source
function and a uniformly bright stellar disk, the unique X-type
solution proves to be a stable attractor to which
time-dependent numerical radiation-hydrodynamical simulations relax.  But in
models where the scattering strength is only modestly reduced,
the topology instead turns nodal, with the associated solution degeneracy
now manifest by intrinsic structure and variability that persist down to the
photospheric wind base. This highlights the potentially
crucial role of diffuse radiation for the dynamics and variability of
line-driven winds, and seriously challenges the use of Sobolev theory
in the transonic wind region. Since such Sobolev-based models are
commonly used in broad applications like stellar evolution and
feedback, this prompts development of new wind models, not only for 
further quantifying the intrinsic variability found here,
but also for computing new theoretical predictions of global 
properties like velocity laws and mass-loss rates.

\end{abstract}


\section{Introduction}
\label{sec:intro}

For massive, hot stars of spectral types OBA, scattering and absorption in spectral 
lines transfer momentum from the star's intense radiation field to the plasma, and so provide the force necessary to
overcome gravity and drive a stellar wind outflow \citep[see][for an extensive 
review]{Puls08}. The first quantitative description of such line-driving was given in the seminal 
paper by \citet{Castor75}: hereafter ``CAK". Like most wind models to date, 
CAK used the so-called Sobolev approximation \citep{Sobolev60} 
 to compute the radiative 
acceleration. This assumes that basic hydrodynamic flow quantities\footnote{Or more 
specifically, occupation number densities and source functions.} are 
constant over a few Sobolev lengths $ \ell_{\rm Sob} = v_{\rm th}/(dv/dr)$ 
(for radial velocity gradient $dv/dr$ and ion thermal speed $v_{\rm th}$), 
allowing then for a \textit{local} treatment of the line radiative transfer.  

Such a Sobolev approach ignores the strong `line deshadowing instability' (LDI)
that occurs on scales near and below the Sobolev length \citep{Owocki84, Owocki88, Feldmeier97, 
Dessart03, Sundqvist13};
this is generally believed to be the origin for extensive 
small-scale structure and `clumping' in the winds from 
hot stars \citep[e.g.,][]{Eversberg98, Puls06, 
Sundqvist10}. 
Indeed, while Sobolev-based models have had 
considerable success in explaining many 
qualitative features of stellar winds from
OBA stars \citep{Pauldrach94, Vink00}, 
recent observational analyses \citep{Najarro11, Sundqvist11,
Surlan13, Cohen14, Sundqvist14, Rauw15} that account 
properly
 for clumping 
systematically infer mass-loss
rates in Galactic O-star winds that are 
factors of at least $\sim2-3$ lower than predicted by 
such Sobolev-based theoretical models.
Such mass-loss reductions  have dramatic consequences for
predictions of stellar evolution and feedback models
\citep[e.g.,][]{Smith14}.

But a further,  arguably related shortcoming of the Sobolev 
approach regards its questionable applicability in regions near and below 
the wind sonic point, where the characteristic scale height for variations 
$H = |  \rho/(d\rho/dr) | \approx | v/(dv/dr) | \la \ell_{\rm Sob} $. \
While early explorations using full co-moving frame transfer solutions 
to compute the radiative acceleration seemed to suggest quite good 
agreement with the Sobolev approach \citep{Pauldrach86}, more 
modern calculations indicate markedly lower radiative acceleration in the 
subsonic region leading up to the wind sonic point (\citealt{Lucy07a, Lucy07b,Lucy10}, 
\citealt{Krticka10}; \citealt{Bouret12}; Sundqvist et al., in prep). Such a reduced 
acceleration in the transonic region can have a significant effect on the wind 
dynamics, perhaps leading to a lower mass-loss rate than predicted by corresponding
Sobolev models, and thus to better agreement with empirically inferred rates.

As analyzed by \citet{Owocki99}, 
the net line-driving in the transonic region involves contributions from both a 
{\em direct} component of radiative momentum flux
{\em absorbed} from the underlying hydrostatic atmosphere,
and a {\em diffuse} component of radiation {\em scattered} within the local line 
resonance\footnote{We emphasize that the ``diffuse'' component here refers to radiation scattering {\em within} the resonance zone for a single, potentially isolated line. This is distinct from the {\em multi-line} scattered radiation that results when two or more distinct  lines have overlapping resonances throughout the global wind.  The latter is commonly modeled with Monte Carlo methods  \citep[e.g.,][]{Vink00},  using Sobolev-based solutions for location of each line resonance, neglecting any net recoil from asymmetries in the radiative escape from within that resonance zone.}.
In the supersonic portions of an idealized, steady-state, unclumped and monotonically accelerated wind, 
application of a Sobolev line-transfer treatment shows that the escape of radiation scattered within the local Sobolev resonance layer attains a fore-aft symmetry, implying then that the 
diffuse component of the line force  {\em vanishes} in a Sobolev approach \citep{Castor74}.
But, as detailed further below, in the transonic region, {\em fore-aft asymmetries} in this escape probability lead to a non-zero diffuse line-force.
 
Nonetheless, because this diffuse term plays little role in driving the bulk of the supersonic wind outflow,  initial dynamical explorations of relaxing the Sobolev approach assumed a {\em pure-absorption} model that completely neglects this diffuse radiation.
Building on the linear stability analysis of \citet{Owocki84}, \citet{Owocki88} used a pure-absorption approach in time-dependent simulations to follow the nonlinear growth of instabilities on scales near and below the Sobolev length, evaluating the direct line-force through non-local, integral calculations for line-absorption of the underlying stellar continuum.

\citet[][hereafter POC]{Poe90} then applied this pure-absorption approach to study the nature of associated steady-state wind solutions.
A major surprise of this analysis was that, instead of the unique transonic solution of the usual X-type critical point, the solutions near the sonic point of such an absorption-line-driven wind exhibit a {\em nodal topology}
\citep{Holzer77}, with both a well-defined, steeper-sloped solution, and a {\em degenerate family} of shallower but still positively sloped solutions (see figure \ref{fig:top}).
Moreover, for realistic values of the ratio between the ion thermal speed and the sound speed (i.e., $v_{\rm th}/a \lesssim 0.3$),  POC showed that physically realistic boundary conditions favor the lower-slope, degenerate solutions, and that in time-dependent simulations this leads to an {\em intrinsic variability} that extends down to the subsonic wind base.

Within the context of steady-state models of the transonic flow,
\citet{Lucy07a, Lucy07b} used non-Sobolev, Monte-Carlo scattering line-transfer in a large number ($> 10,000$) of spectral lines to compute the total (direct + diffuse) line force.
In deriving  solutions for the associated mass flux through the sonic point, the analytic parameterization for this numerical, Monte Carlo line force was assumed to depend not on the velocity gradient, as it does in Sobolev-based models, but solely on the flow velocity itself, without any explicit dependence on the spatial position.
As discussed further below (see the end of \S\ref{sec:sonictop}), such an assumed pure-velocity scaling
recovers the X-type topology, with one unique positive slope solution velocity solution through the sonic point.

To incorporporate scattering effects in time-dependent dynamical simulations, 
\citet{Owocki91} 
\citep[see also][hereafter OP96]{Owocki96}
introduced a ``smooth source function'' (SSF) method;
this accounts for non-Sobolev asymmetries in the frequency and angle-dependent escape probabilities, but assumes that the scattering source function -- as a frequency and angle {\em averaged} quantity -- remains smooth and so effectively constant through the resonance zone.
 In this SSF approach, asymmetries in the escape near the sonic point lead to a net positive diffuse line-force that tends to compensate for the photospheric-line-shadowing reduction in the direct component, giving then a total line-force that tends to be quite close to the CAK/Sobolev value.
Remarkably, for the simple case of an optically thin scattering source function without any limb darkening, the base of
time-dependent SSF simulations relax to a steady state that matches quite well the standard CAK/Sobolev solutions, with now a completely {\em stable} transonic wind base \citep{Owocki91, Owocki99}.

However, more recent simulations by \citet[][hereafter SO13]{Sundqvist13} that account for photospheric {\em limb darkening} now again show intrinsic base variability and associated structure in near photospheric layers, in accordance with empirical results derived 
from a multitude of O-star observations in various wavebands \citep[e.g.,][]{Eversberg98, Puls06, Najarro11, Cohen11, Sundqvist11, Bouret12, Cohen14}. This suggests that even a relatively minor associated reduction in scattering strength can have potentially important implications for formation of wind structure, and thus also for the empirical derivation of wind mass loss rates. This motivates here a renewed examination of the dynamics of line driving from the transonic wind base.

In particular, to clarify and quantify the subtle, but potentially crucial role of diffuse, scattered radiation in the initiation and stability of a line-driven wind through this transonic region, \S\ref{sec:sseq} extends the POC pure-absorption solution topology analysis to account for scattering effects within the SSF formalism.
A key result is that, below some minimum level of scattering, the transonic solution topology transitions from  the well-defined X-type to the degenerate nodal type.
Comparison with time-dependent simulations in \S\ref{sec:ssfsim} shows a remarkably close correspondence between the conditions for this steady-state topology and the transition from stable relaxation to
intrinsic base variability.
The discussion in \S\ref{sec:discussion} interprets these results in the broader context of modeling line-driven mass loss,
and outlines some directions for future work.

\section{Steady-State critical point analysis}
\label{sec:sseq}

\subsection{Direct and Diffuse Components of the SSF Line-Force}
\label{sec:ssf}

For an isothermal, spherically symmetric,  line-driven stellar wind,
the steady-state equation of motion for variation of radial velocity $v$ with radius $r$ can be written as
\beq
\left ( v - \frac{a^2}{v} \right ) \, \frac{dv}{dr} =
- \frac{GM_{\rm eff}}{r^2}
+ g_{\rm lines} + \frac{2 a^2}{r}
\, ,
\label{eq:eom}
\eeq
where the terms with isothermal  sound speed $a$ account for the effects of gas pressure.
Here the gravity $GM_{\rm eff}/r^2$ scales with an effective stellar mass $M_{\rm eff} \equiv M (1-\Gamma_{\rm e})$, reduced from the outward force of electron scattering by a fraction set by the Eddington parameter $\Gamma_{\rm e} = \kappa_{\rm e} L/(4 \pi G M c)$, with
$\kappa_{\rm e}$ the electron scattering opacity, $M$ and $L$ the stellar mass and luminosity, 
$G$ the gravitation constant, and $c$ the speed  of light.
Since the radiative flux and thus line acceleration $g_{\rm lines}$ have a similar inverse-square dependence on radius $r$, it is convenient to also scale this by the effective gravity,
\beq
\Gamma_{\rm lines} 
\equiv \frac{g_{\rm lines}}{GM_{\rm eff}/r^2} 
= \Gamma_{\rm dir} + \Gamma_{\rm diff}
\, ,
\label{eq:Gamrad}
\eeq
where the latter equality identifies this total line acceleration as given by the sum of distinct direct and diffuse components.
The steady-state topology analysis in POC included only the direct component, but we now wish to extend the pure-absorption treatment  there to account for scattering terms within the SSF formalism (\citealt{Owocki91}; OP96).

Equations (65) and (67) of OP96 provide relevant expressions for these direct and diffuse forces in terms of integrals over  thermally scaled frequency displacement from line center $x $ ($\equiv (\nu - \nu_o)/\Delta \nu_{\rm D}$, where $\Delta \nu_{\rm D} = \nu_o v_{\rm th}/c$ is the Doppler width associated with ion thermal speed $v_{\rm th}$),
 and over ray index $y$ ($ \equiv (p/R)^2$, where $p$ is the ray impact parameter on the projected stellar disk of radius $R$).
Following POC, we here compute these integrals in terms of associated quadrature sums over discrete grids in $x$ and $y$.
For the direct, absorption component of gravitationally scaled line-acceleration, we thus write
\beq
\Gamma_{\rm dir} =  \Gamma_{\rm thin}  \sum_{x,y} w_{xy}  \, 
 \phi \left( x-\mu_y u \right )  t_+^{-\alpha} 
\, ,
\label{eq:gamdir}
\eeq
with $w_{xy} = w_x w_y$ a suitable set of quadrature weights in $x$ and $y$, and 
with $\alpha$ the power-law index in the assumed line-distribution function 
(see, e.g., equation 52 in OP96). Here
$\phi$ is the line profile function (taken to be a Gaussian),
with $u \equiv v/v_{\rm th}$ the thermally scaled radial flow speed,
projected along the ray by the direction cosine
\beq
\mu_y (r) \equiv \sqrt{1 - y \frac{R^2}{r^2}}
\, .
\label{eq:muydef}
\eeq
As in OP96, the line acceleration here is normalized by the optically thin form for a line of fixed opacity $\kappa_{\rm o}$,
\beq
\Gamma_{\rm thin} 
= \frac{\kappa_{\rm o} v_{\rm th} L}{4 \pi  GM_{\rm eff} c^2}
= \frac{\kappa_{\rm o} v_{\rm th}}{\kappa_{\rm e} c}
 \frac{ \Gamma_{\rm e} }{1- \Gamma_{\rm e}}
\, ,
\label{eq:Gamthin}
\eeq
wherein\footnote{
In the notation of \citet{Gayley95}, the line normalization here is given by 
$ \kappa_{\rm o} v_{\rm th} /\kappa_{\rm e} c =  \left [ {\bar Q} Q_{\rm o}^{- \alpha}/\Gamma(\alpha) \right ]^{1/(1-\alpha)}$,
where $\Gamma(\alpha)$ is the complete Gamma function. 
The quoted numerical value then follows from ${\bar Q} \approx Q_{\rm o} \approx 2000$ and $\alpha \approx 0.65$
\citep{Gayley95, Puls00}. Note further that in this implementation of the OP96 notation, the frequency weights $w_x$ here must sum to $\Gamma (\alpha)$.}
the frequency-integrated line strength $ \kappa_{\rm o} v_{\rm th} /\kappa_{\rm e} c \approx 1000 $
for a typical Galactic O-star \citep{Puls00}. 

The $t_+$ term in equation (\ref{eq:gamdir}) represents an associated outwardly ($+$) integrated line optical depth, 
as given, e.g., by equation (66) of OP96.
Within the SSF formalism, the {\em diffuse} component of the line acceleration can be written in terms of the 
difference between this term and the {\em inwardly} integrated ($-$) optical depth $t_-$, defined by OP96 equation (68), 
\beq
\Gamma_{\rm diff} = s(r) \, \Gamma_{\rm thin}  \sum_{x,y} w_{xy}  \, 
 \phi \left( x-\mu_y u \right )  \left [  t_-^{-\alpha} - t_+^{-\alpha}  \right ]
\, .
\label{eq:gamdiff}
\eeq
The $t_{\pm}$ integral forms (66) and (68) in OP96 are obtained by applying appropriate boundary conditions to radial integrations of the differential form for these outward/inward optical depths,
\beq
\frac{dt_\pm (\pm x,y,r)}{dr} =
\pm \frac{\kappa_o \rho(r) \phi (x - \mu_y (r) u (r))}{\mu_y (r)}
\, ,
\label{eq:dtpmdr}
\eeq
a relation that will prove useful in the topology analysis below.
The scattering factor $s$ represents the relative strength of the scattering source function to the radial flux.
Following OP96, we write this as 
\beq
s (r) = \frac{0.5}{1+\mu_1(r)} \, \frac{S}{S_{\rm thin}}, 
\, 
\label{eq:sdef}
\eeq
with source function $S$ and $S = S_{\rm thin} = (1-\mu_1(r))/2$ for the simplest case 
of an optically thin source function from a star without any limb darkening. 

\subsection{Sonic Point Topology in SSF Model}
\label{sec:sonictop}

These SSF forms for the line-acceleration provide the basis for generalizing the pure-absorption solution topology analysis of POC to now account for the effects of the diffuse, scattered radiation.
In analogy to equation (1) of POC, let us rewrite the equation of motion (\ref{eq:eom}) in the form
\beq
\frac{r^2}{GM_{\rm eff}} \, 
\frac{dv}{dr} 
= \frac{\Gamma_{\rm lines} - 1 + 2a^2 r/GM_{\rm eff}}{v-a^2/v} \equiv \frac{N}{D}
\, .
\label{eq:NbD}
\eeq
As in POC, our analysis here gives the line acceleration in terms of both the local radius $r$ and local speed $v$, allowing one then to derive the topology of solutions near the critical, sonic point, where
\beq
N(r_{\rm c}) = D(r_{\rm c}) \equiv 0 
~ => ~ 
v_{\rm c} \equiv  v(r_{\rm c}) \equiv a \equiv u_{\rm c} v_{\rm th}
\, ,
\eeq
with associated definitions
\beq
{v_{\rm c}'} \equiv \left . \frac{dv}{dr} \right |_{\rm c} 
~ ; ~
g_{\rm c} \equiv \frac{GM_{\rm eff}}{r_{\rm c}^2}
~ ; ~
\phi_{\rm c} (x) \equiv \phi (x - \mu_{yc} u_{\rm c} )
\, .
\label{eq:vcgcdef}
\eeq
Using L'Hopital's rule $[N/D]_{\rm c} = [dN/dr]_{\rm c}/[dD/dr]_{\rm c}$, we obtain
\beqa
{v_{\rm c}'}^2 &= & \frac{g_{\rm c}}{2} \, \left . \frac{dN}{dr} \right |_{\rm c}
\nonumber
\\
&=& 
\frac{a^2}{r_{\rm c}^2}
+ \frac{g_{\rm c}}{2} \, \left . \frac{d\Gamma_{\rm lines}}{dr} \right |_{\rm c}
\nonumber
\\
&=&
\frac{a^2}{r_{\rm c}^2}
+  \frac{g_{\rm c}}{2} \, \left  [ \left . \frac{\partial\Gamma_{\rm lines}}{\partial r} \right |_{\rm c} 
+ v_{\rm c}'  \left . \frac{\partial\Gamma_{\rm lines}}{\partial v} \right |_{\rm c} 
\right ]
\nonumber
\\
&\equiv & A + B v_{\rm c}'
\, .
\label{eq:vcpdef}
\eeqa
Applying the above SSF model for the total radiative acceleration, along with equation (\ref{eq:dtpmdr}) to evaluate the radial derivatives of the optical depths $t_{\pm}$,  the coefficients $A$ and $B$ now take the explicit forms,

%
\beqa
A   - \frac{a^2}{r_{\rm c}^2}  &\equiv&  \frac{g_{\rm c}}{2} \, \left . \frac{\partial\Gamma_{\rm lines}}{\partial r} \right |_{\rm c} 
\label{eq:Adef}
\\
&=& \frac{g_{\rm c}}{2} \Gamma_{\rm thin} \sum_{x,y} w_{xy} 
\bigl \{ \bigr .
\nonumber
\\
&-& 
\alpha \frac{\kappa_o \rho_{\rm c}}{\mu_{yc}} \, \phi_{\rm c} ^2
 \left [ t_{\rm c+}^{-1-\alpha} - s_{\rm c} ( t_{\rm c+}^{-1-\alpha}  +  t_{\rm c-}^{-1-\alpha}) \right ]
 \nonumber
\\
&-& s'_{\rm c} 
 \, \phi_{\rm c}  \left [ t_{\rm c+}^{-\alpha}  -  t_{\rm c-}^{-\alpha} \right ]
 \nonumber
\\
&+& 
\left . 
2 \mu'_{yc} \,  u_{\rm c} \, (x - \mu_{yc} u_{\rm c} ) \phi_{\rm c}  
\left [ t_{\rm c+}^{-\alpha} - s_{\rm c} ( t_{\rm c+}^{-\alpha}  -  t_{\rm c-}^{-\alpha}) \right ]
\right \}
 \nonumber
\\
&& 
\label{eq:pgradpr}
\eeqa
and
\beqa
B &\equiv& \frac{g_{\rm c}}{2} \, \left . \frac{\partial\Gamma_{\rm lines}}{\partial v} \right |_{\rm c} 
\nonumber
\\
&=&
\frac{g_{\rm thin,c}}{v_{\rm th}} 
\sum_{x,y} w_{xy} 
\nonumber
\\
&&
 \mu_{yc} \,  (x - \mu_{yc} u_{\rm c} ) \phi_{\rm c}  
\left [ t_{\rm c+}^{-\alpha} - s_{\rm c} ( t_{\rm c+}^{-\alpha}  -  t_{\rm c-}^{-\alpha}) \right ]
\, .
\label{eq:Bdef}
\eeqa
For pure-absorption ($s_{\rm c} = s'_{\rm c} = 0$) along a radial ray ($y=0; \mu_y = \mu_0 = 1$; $\mu'_y=0$), 
the last two terms in (\ref{eq:pgradpr}) vanish, along with the $s_{\rm c}$ elements of the first term, giving then just the POC expression for $A$ (cf. their equation 12).
Likewise, the vanishing of the $s_{\rm c}$ terms in (\ref{eq:Bdef}) means that this also recovers the POC expression for $B$
(cf. their equation 13).

From the quadratic equation, we see that solutions for the critical point slope take the form
\beq
v_{\rm c}' (\pm) = \frac{B \pm \sqrt{B^2 + 4 A}}{2}
\, .
\label{eq:quad}
\eeq
As shown by POC, for the pure-absorption case, $A<0$ and $B>0$.
If $B^2 + 4A > 0$, the critical point is then {\em nodal} type, with {\em two positive} slope solutions, 
while for $B^2 + 4A < 0$), it is {\em focal} type, with {\em no real} solutions for $v_{\rm c}'$.

\begin{figure}
\begin{center}
\includegraphics[scale=0.32]{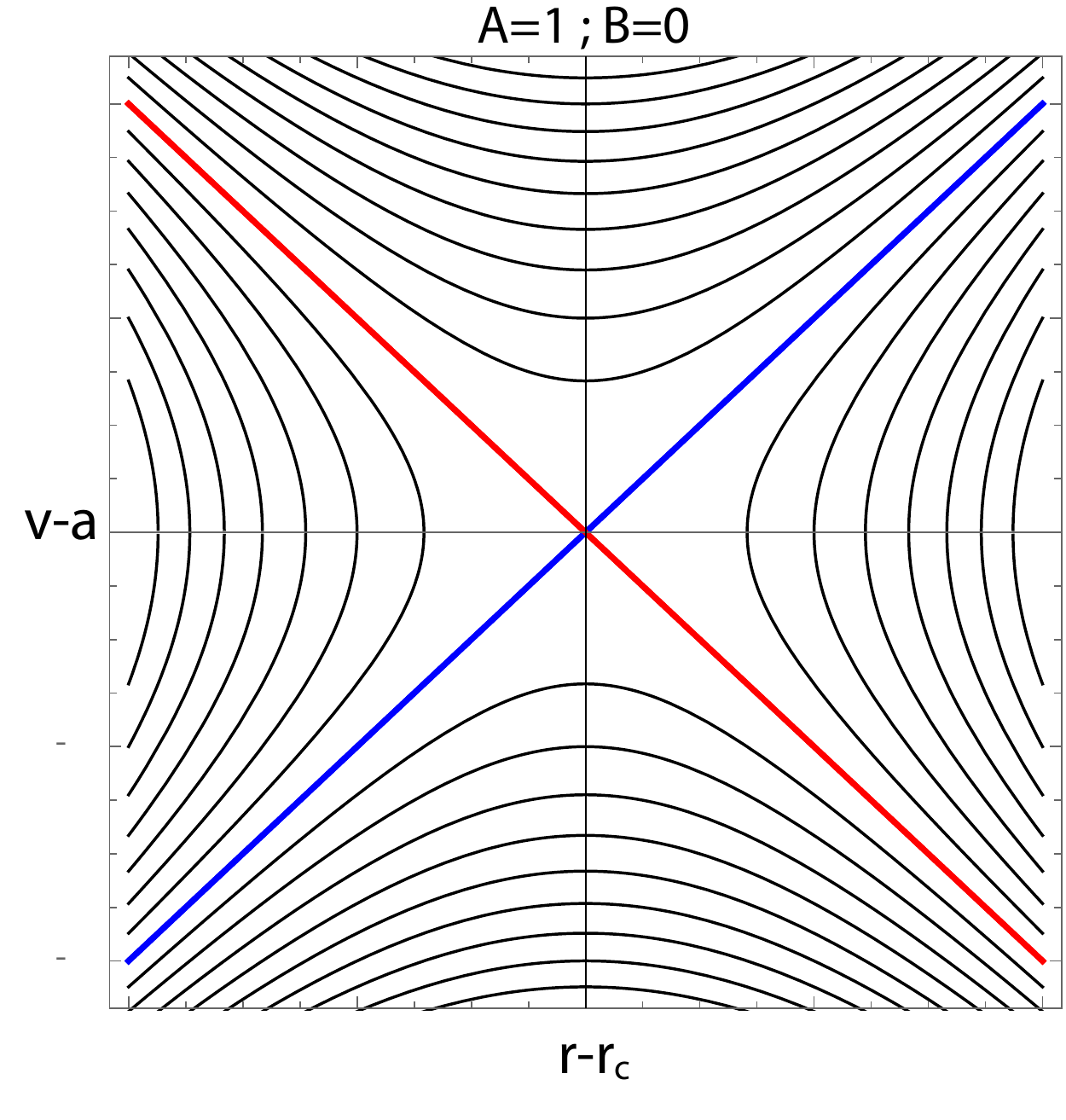}
\includegraphics[scale=0.32]{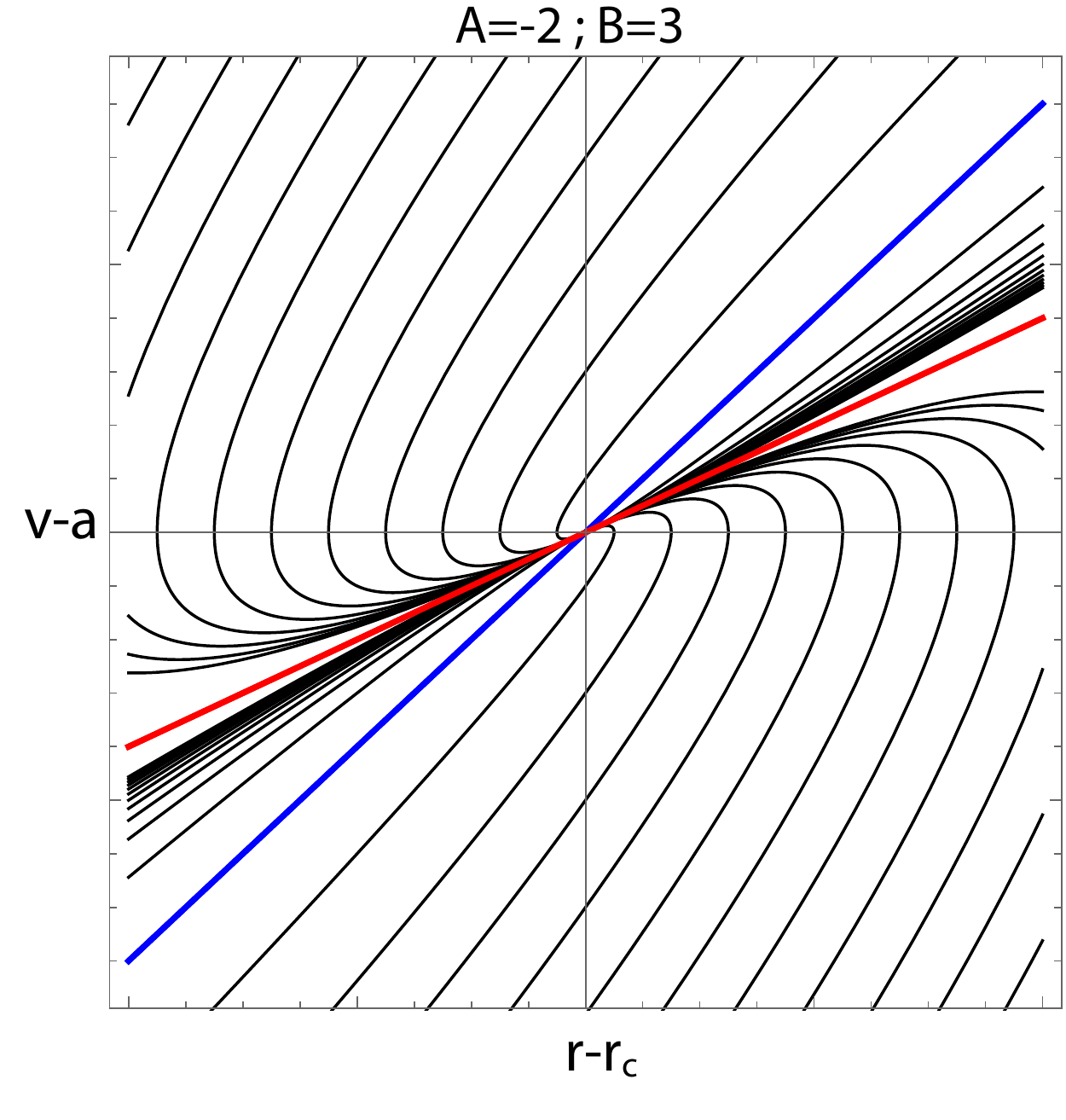}
\caption{Illustrations of solution topologies near the critical (sonic) point. 
The left panel shows an X-type case with $A=1$, $B=0$ and so $v_{\rm c\pm}' = \pm 1$ (blue/red lines).
The right panel shows a nodal case with $A=-2$, $B=3$, and so $v_{\rm c+}' = +2$  (blue line) and $v_{\rm c-}' = +1$ (red line).
}
\label{fig:top}
\end{center}
\end{figure}

But the addition of the scattering terms now makes it possible, for sufficiently large $s_{\rm c}$, to have $A>0$, thus implying a saddle or X-type critical point, with two real solutions of $v_{\rm c}'$ that have {\em opposite } signs.
Figure \ref{fig:top} illustrates the distinct differences between an X-type topology (left) vs.\ a nodal topology (right).

For nonzero but modest scattering factor $s_{\rm c}$, the first two terms within the sum in (\ref{eq:pgradpr}) are nominally negative, while the final term is positive.
 But the magnitude of the first term is of order $ dg_{\rm lines}/dr \approx g_{\rm c}/H $, where $H = a^2/g_{\rm c}$ is the atmospheric scale height; in comparison, the latter two terms are typically a factor $H/r_{\rm c} \sim a^2/g_{\rm c} r_{\rm c}  < 0.01$ smaller.
Moreover, the first term is an even bigger factor $r_{\rm c}^2/H^2 > 10^4$ larger than the gas pressure term $a^2/r_{\rm c}^2$.

As such, we may generally neglect all these smaller terms, and so approximate the quantity $A$ by just keeping the first term. 
Defining
\beq
a_\pm \equiv 
 \sum_{y} \frac{w_{y} }{\mu_{yc}} \, \sum_{x} w_x
\frac{\phi_{\rm c}^2}{ t_{\rm c\pm}^{1+\alpha} }
\, ,
\label{eq:apm}
\eeq
we can then write $A$ in the form
\beq
A  \approx  
- \frac{g_{\rm c}}{2} \, \Gamma_{\rm thin}  \, 
\alpha \kappa_o \rho_{\rm c} \, 
~ f_{\rm top} \,  a_+ 
 \, ,
 \label{eq:Aapprox}
\eeq
where we have now defined a `topology function',
\beq
f_{\rm top} \equiv 
1 - s_{\rm c} 
 \left ( 1 + 
 \frac{a_{-}}{a_{+}} 
 \right )
 \, . 
\label{eq:ftop}
\eeq
For the pure-absorption case, with $s_c=0$, we have $f_{\rm top} = 1$ and so $A<0$, giving then a nodal or focal solution topology. To obtain $A>0$, and thus X-type solutions, $f_{\rm top}$ 
must become negative, which requires some combination of large values for $s_{\rm c}$ and/or for the ratio 
$a_-/a_+$.

Finally, within this context, let us also consider the implied topology for the non-Sobolev, Monte-Carlo scattering models of \citet{Lucy07a, Lucy07b}.
Since these parametrize the line force as purely a function of velocity 
(thus with nonzero  $B \sim \partial \Gamma_{\rm lines}/\partial v$)
but with no explicit spatial dependence (and thus vanishing $ \partial \Gamma_{\rm lines}/\partial r$),
they have $A = a^2/r_{\rm c}^2 > 0$, implying then an X-type solution topology, 
with one unique positive slope solution.
Moreover, since $B \sim g_c/a \sim a/H \gg a/r_{\rm c}$, this positive slope is quite steep, $v_{\rm c}' (+) \sim B \sim a/H$, while the negative slope is quite shallow, $- v_{\rm c}' (-) \sim  A/B \sim (H/r_{\rm c})(a/r_{\rm c})$.
While this assumed line-force parametrization thus gives a unique, non-degenerate solution, further work is needed to clarify the physical justification for neglecting any explicit spatial dependence in parametrizing the Monte Carlo line force.

\subsection{Sobolev optical depth for forward/backward streams, $\tau_{\pm}$}
\label{sec:tausobpm}

Appendix A details a formal second-order Sobolev analysis for estimating the expected values of  
$a_\pm $ and thus $A$.
To build insight, let us consider here a simpler, more heuristic analysis for this.
Within the Sobolev approximation, the optical depth integrals scale in proportion to a locally defined Sobolev optical depth, $t \sim \tau \equiv \kappa_o \rho \ell$, where $\ell \equiv v_{\rm th}/v'$ is the Sobolev length.
But for the critical-point forward/backward streaming optical depths $t_{\rm c \pm}$, the relevant evaluations center on regions that extend roughly a Sobolev length below/above  the sonic point;
in terms of sonic point values of the Sobolev length  $\ell_{\rm c} = v_{\rm th}/v_{\rm c}'$ and optical depth $\tau_{\rm c} = \kappa_o \rho_{\rm c} \ell_{\rm c}$,
the associated forward/backward streaming Sobolev optical depths scale as 
\beq
\frac{\tau_{\rm c \pm}}{\tau_{\rm c}}
=
1  \mp  \frac{\ell_{\rm c}}{2} \frac{\tau_{\rm c}'}{\tau_{\rm c} }
\approx 
1  \mp  \frac{\ell_{\rm c}}{2} \frac{\rho_{\rm c}'}{\rho_{\rm c} }
\approx 
1  \pm \frac{\ell_{\rm c}}{2} \frac{v_{\rm c}'}{a}
= 
1  \pm \frac{v_{\rm th}}{2 a}
\, ,
\label{eq:taupm}
\eeq
where $\tau_{\rm c}'$ represents the local radial gradient of $\tau_{\rm c}$.
Here the second approximation assumes that the velocity gradient is itself nearly constant (i.e., that
$v_{\rm c}'' \approx 0$), while the latter forms use the constancy of mass flux $\rho v$ for a steady, nearly planar outflow.

Since $t_{\rm c \pm} \sim \tau_{\rm c \pm}$, we then see that 
\beq
\frac{a_-}{a_+} \approx \left ( \frac{\tau_{\rm c+}}{\tau_{\rm c-}} \right )^{1+\alpha}
\approx  1 + (1+\alpha) \frac{ v_{\rm th}}{a }
\, ,
\label{eq:ampapprox}
\eeq
where the latter approximation applies first order expansion in the small parameter $v_{\rm th}/a < 1 $.
Applying this in (\ref{eq:ftop}), we can estimate a minimum value for the scattering constant to obtain $f_{\rm top} < 0$ and thus $A> 0$ for an X-type critical solution,
\beq
s_{\rm c} > s_{\rm X} 
\equiv \frac{1}{1+ 
\frac{a_-}{a_+} }
\approx \frac{1}{2+
 \frac{ (1+\alpha) v_{\rm th}}{a }
 }
\, .
\label{eq:sX}
\eeq

In the strict Sobolev limit $v_{\rm th} \rightarrow 0$,
obtaining an X-type neutral point would require a sonic region scattering factor $s_{\rm c} > s_{\rm X} = 1/2$. 
For a simple optically thin scattering source-function, 
a surface intensity that is either constant or limb {\em darkened} has $s(r) \le 1/2$ for all $r \ge R$
(see equation \ref{eq:sdef}).
Such a strict Sobolev limit model would thus have to have photospheric {\em limb brightening} to achieve the $A>0$ needed for an X-type solution topology.

But for a small but nonzero thermal speed ratio $v_{\rm th}/a < 1$, equation (\ref{eq:sX}) gives $s_X < 1/2$; 
for example, for $\alpha=0.65$ and $ v_{\rm th}/a \approx 0.28$, we find $s_{\rm X} \approx 0.4$, making it easier to get $A>0$ and so a stable, X-type critical solution.
This can thus help explain why the base of SSF models without limb darkening relax to a stable steady state \citep{Owocki91, Owocki99}. However, when limb darkening is included, the lower value of $s_{\rm c}$ can again lead to $A<0$, and so a nodal topology, with its associated intrinsic variation at the wind base, as found in the SSF simulations by SO13.

Within the more formal second-order Sobolev analysis of Appendix A, figure \ref{fig:apm-int} illustrates the asymmetries in $t_{\rm c\pm}$ that lead to $a_- > a_+$, while figure  \ref{fig:apmvsvth} plots the resulting variations of $a_-/a_+$ and $s_X$ with $v_{\rm th}/a$.

\section{Comparison with time-dependent SSF Simulations}
\label{sec:ssfsim}

Let us now compare and interpret the above analytic results in the context of 
full numerical time-dependent hydrodynamic wind simulations based on the SSF model
for line-driving.

\subsection{Basic Model Description} 
\label{sec:basic}

The simulations here use the numerical hydrodynamics code\footnote{The VH-1 hydrodynamics computer-code package has been developed 
by J. Blondin and collaborators, and is available for download  at: 
\url{http://wonka.physics.ncsu.edu/pub/VH-1/}} \mbox{VH-1} to evolve the conservation equations of mass and 
momentum for a spherically symmetric and 
isothermal outflow. As in previous SSF simulations \citep[e.g.,][SO13]{Owocki99}, 
we implement radiation line-driving following the method described in \S\ref{sec:ssf}. 
All results presented here adopt the same stellar and wind parameters as in SO13, given 
by their Table 1, which are typical for an O-star in the Galaxy.
For the models in \S\ref{sec:sthin}. and \S\ref{sec:limbdark} we fix  the ratio between ion thermal speed and isothermal sound speed to $v_{\rm th}/a = 0.28$, representative of driving by carbon, 
oxygen, and nitrogen atoms (as in SO13; see also discussion in POC);
in \S\ref{sec:vthba}  we explore different fixed values of $v_{\rm th}/a$ to study its influence on the stability of the wind base.  

Each simulation evolves from a smooth, CAK-like initial condition, computed by relaxing to a steady state a time-dependent simulation that uses a CAK/Sobolev form for the line-force.
To prevent artificial structure due to numerical truncation errors
we use a fixed evolution time-step of 2.5~s, rather than 
setting this step to a fixed fraction of the courant time (see discussion in POC). 
As in previous work, the lower boundary at the assumed stellar 
surface fixes the density to a value $\sim 5-10$ times that at 
the sonic point. Since we are interested here in topology and intrinsic wind 
variability, we do not introduce any explicit base perturbations 
(see SO13, for examples of SSF
simulations including such explicit perturbations); this means all structure 
seen in the models here is truly  intrinsic to the line-driven wind. 
The spatial grid uses 500 discrete radial mesh-points out to $r = 1.5R_\star$, 
 with a step-size that increases linearly with radius,
to increase the resolution in the transonic region critical for 
the analysis here. Finally, all simulations use a single ray 
coordinate at $y=0.5$ to approximate the effects on the line-force from 
the finite stellar disc; tests have shown that such a single ray 
typically gives a radiative acceleration that deviates $\la 10$\% 
from that computed from full angle integration (see SO13; their 
Figure A.1). 

\subsection{Simulation Results} 
\label{sec:simresults}

\begin{figure*}
    \vspace{1cm}
  \begin{minipage}{8.0cm}
            \includegraphics[angle=90,width=8.0cm]{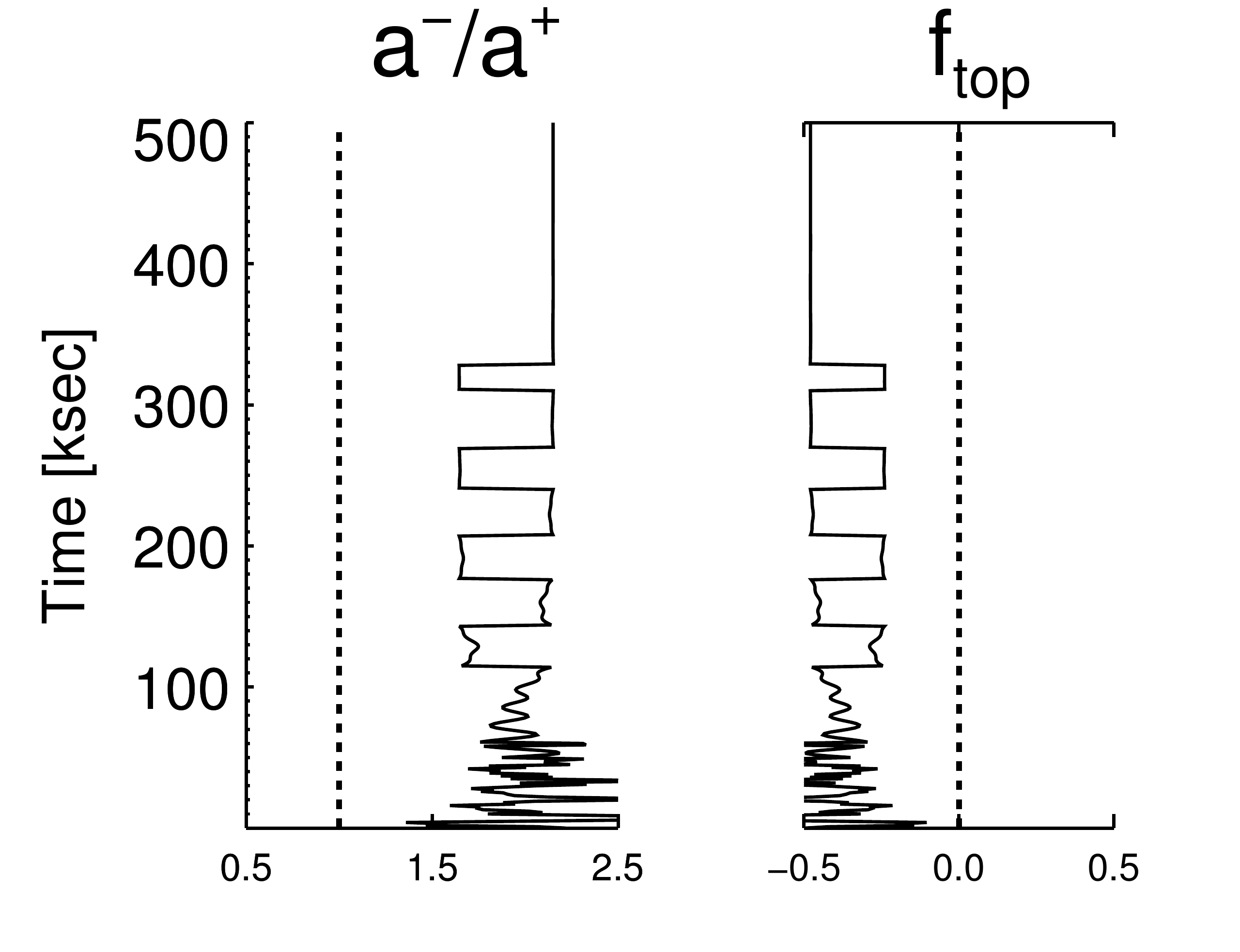}
    \centering
   \end{minipage}
    \begin{minipage}{8.0cm}
         \includegraphics[angle=90,width=8.0cm]{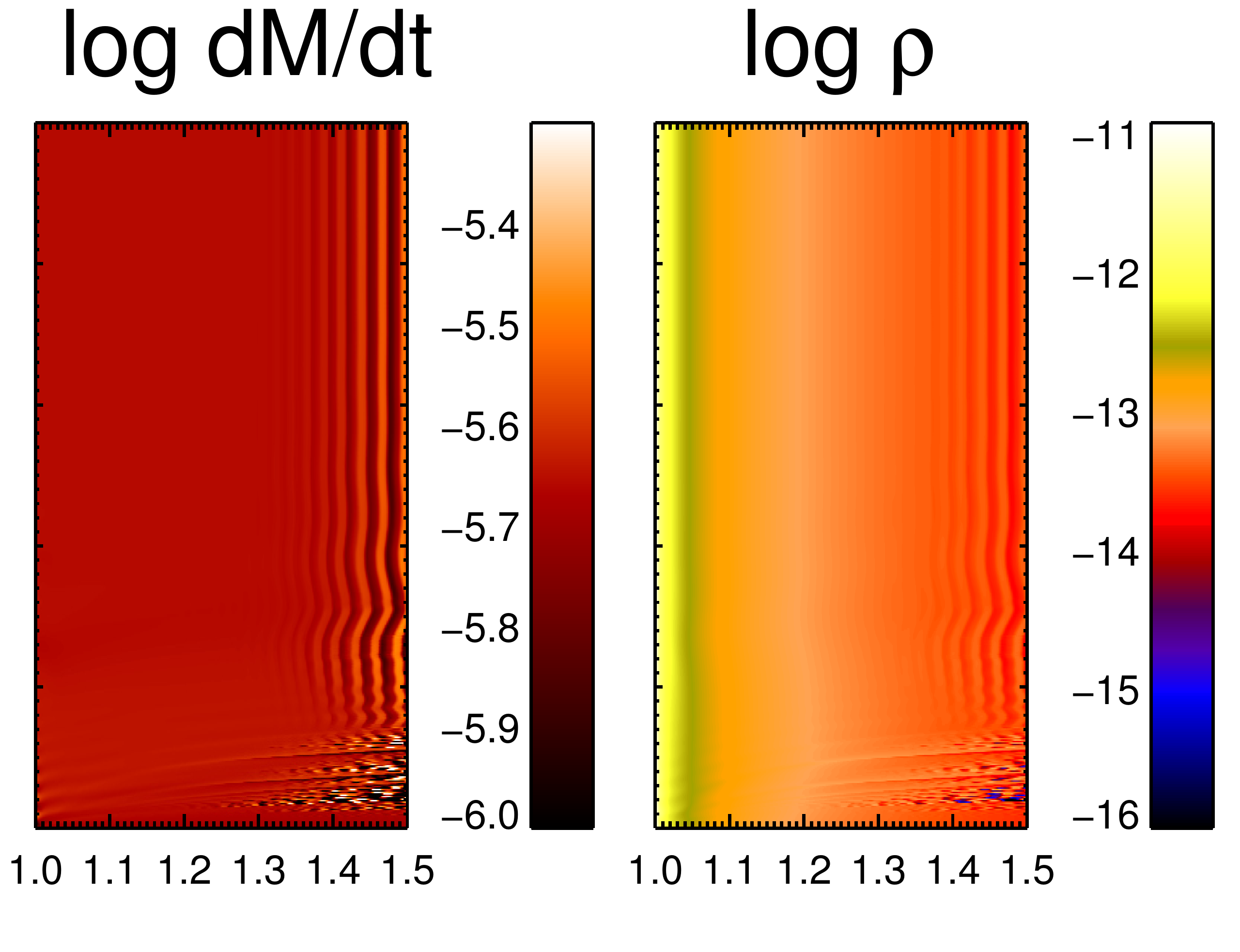}
            \centering
   \end{minipage} 
     \begin{minipage}{8.0cm}
      \includegraphics[angle=90,width=8.0cm]{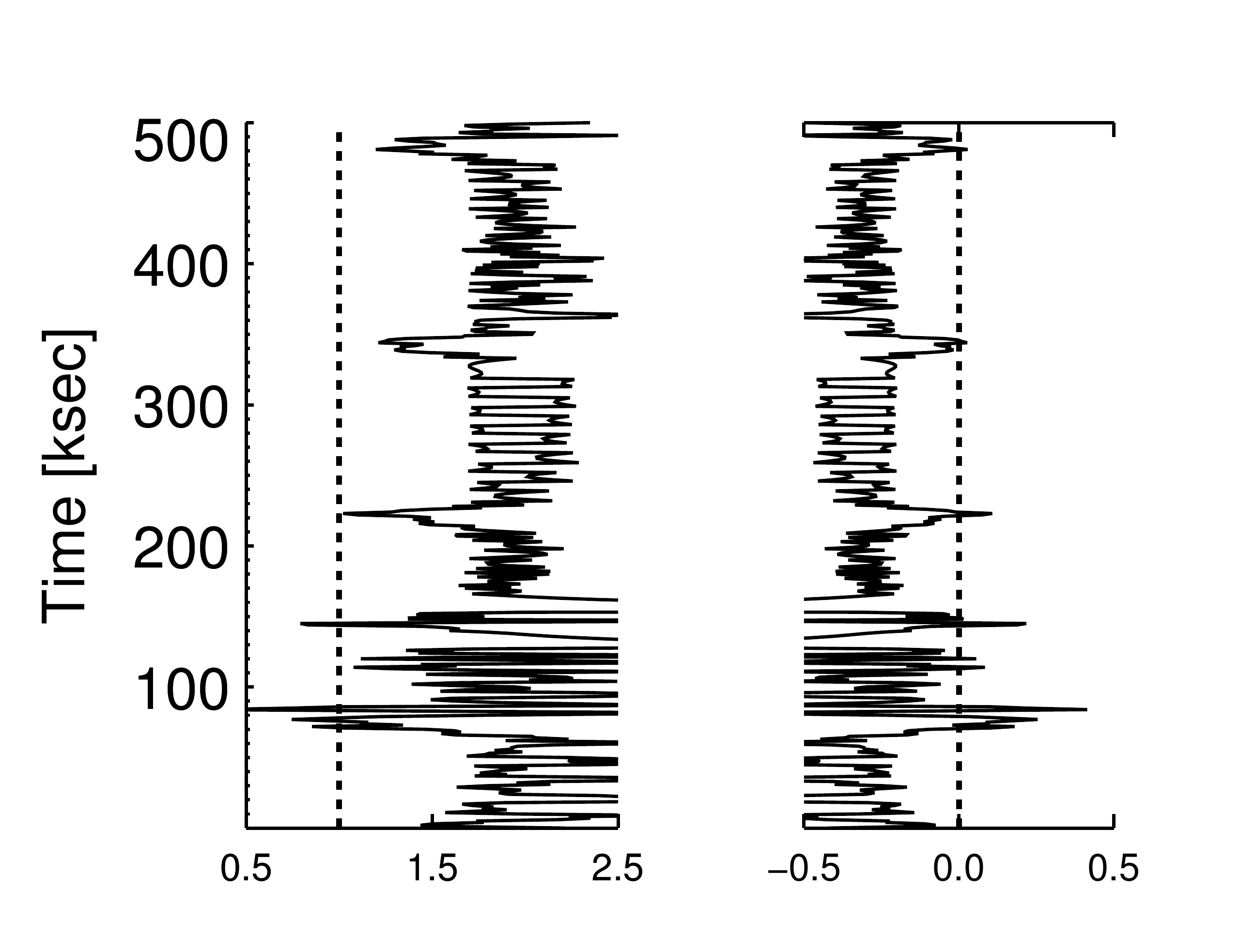}
    \centering
   \end{minipage}
    \begin{minipage}{8.0cm}
      \includegraphics[angle=90,width=8.0cm]{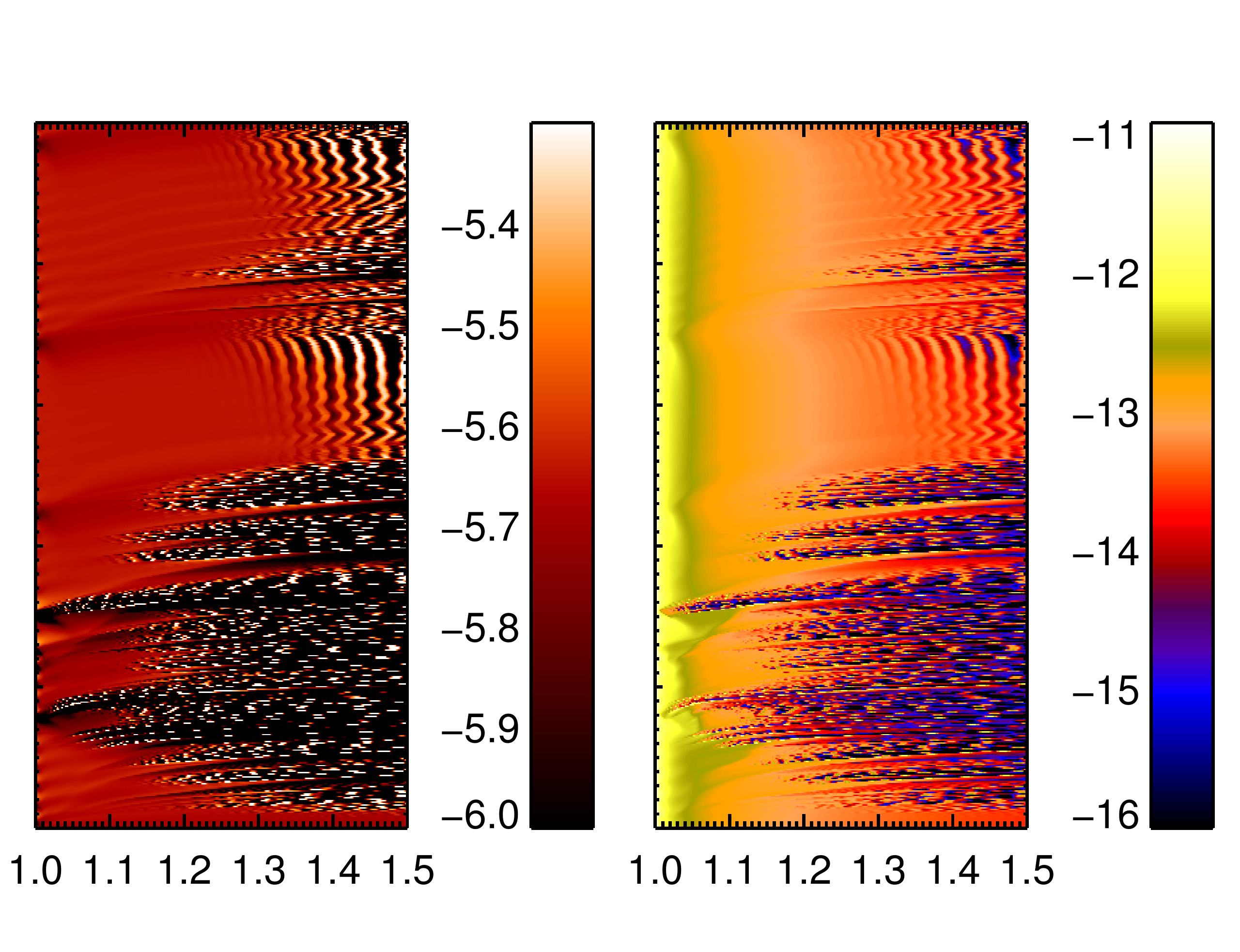}
            \centering
   \end{minipage} 
       \begin{minipage}{8.0cm}
    \includegraphics[angle=90,width=8.0cm]{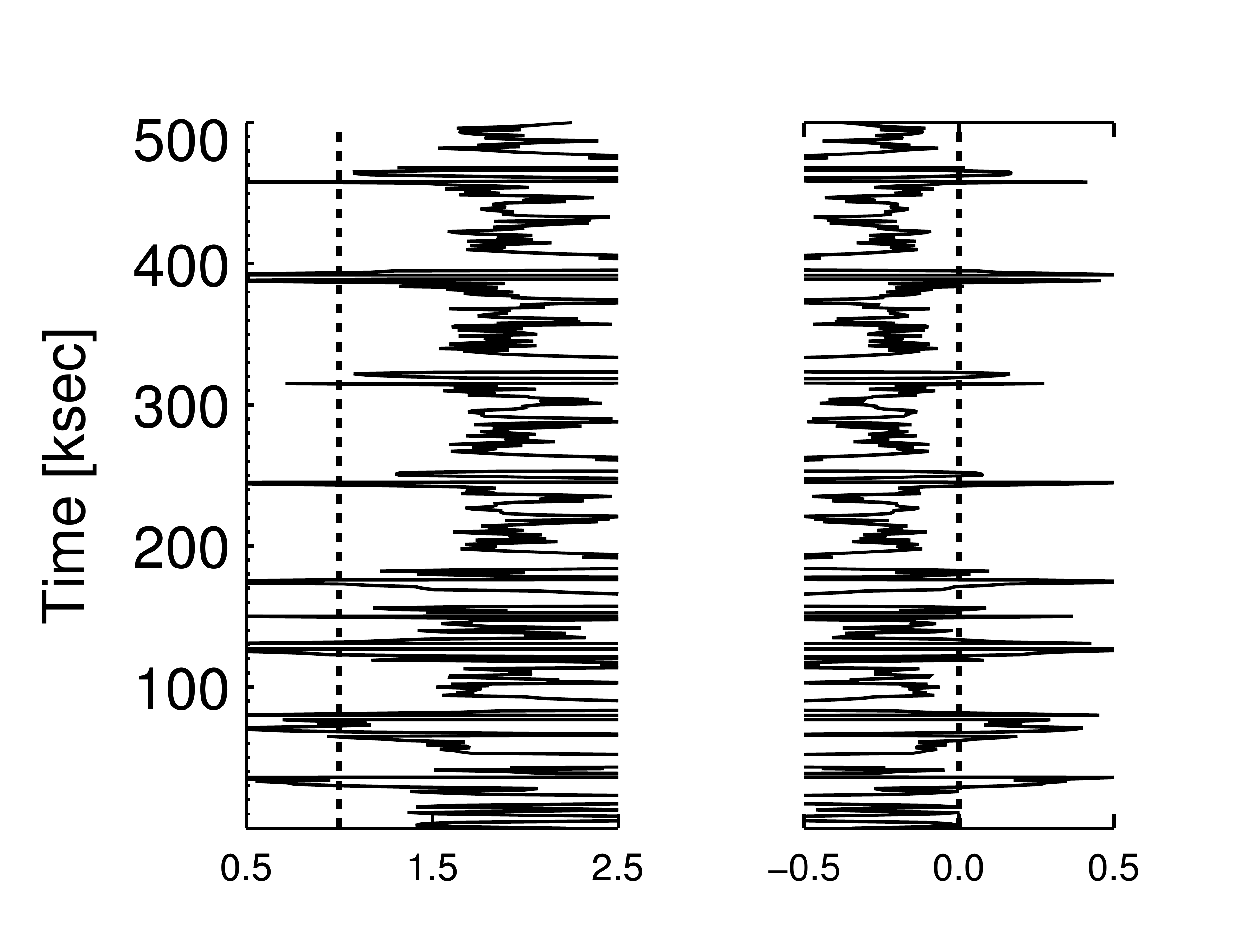}
    \centering
   \end{minipage}
    \begin{minipage}{8.0cm}
      \includegraphics[angle=90,width=8.0cm]{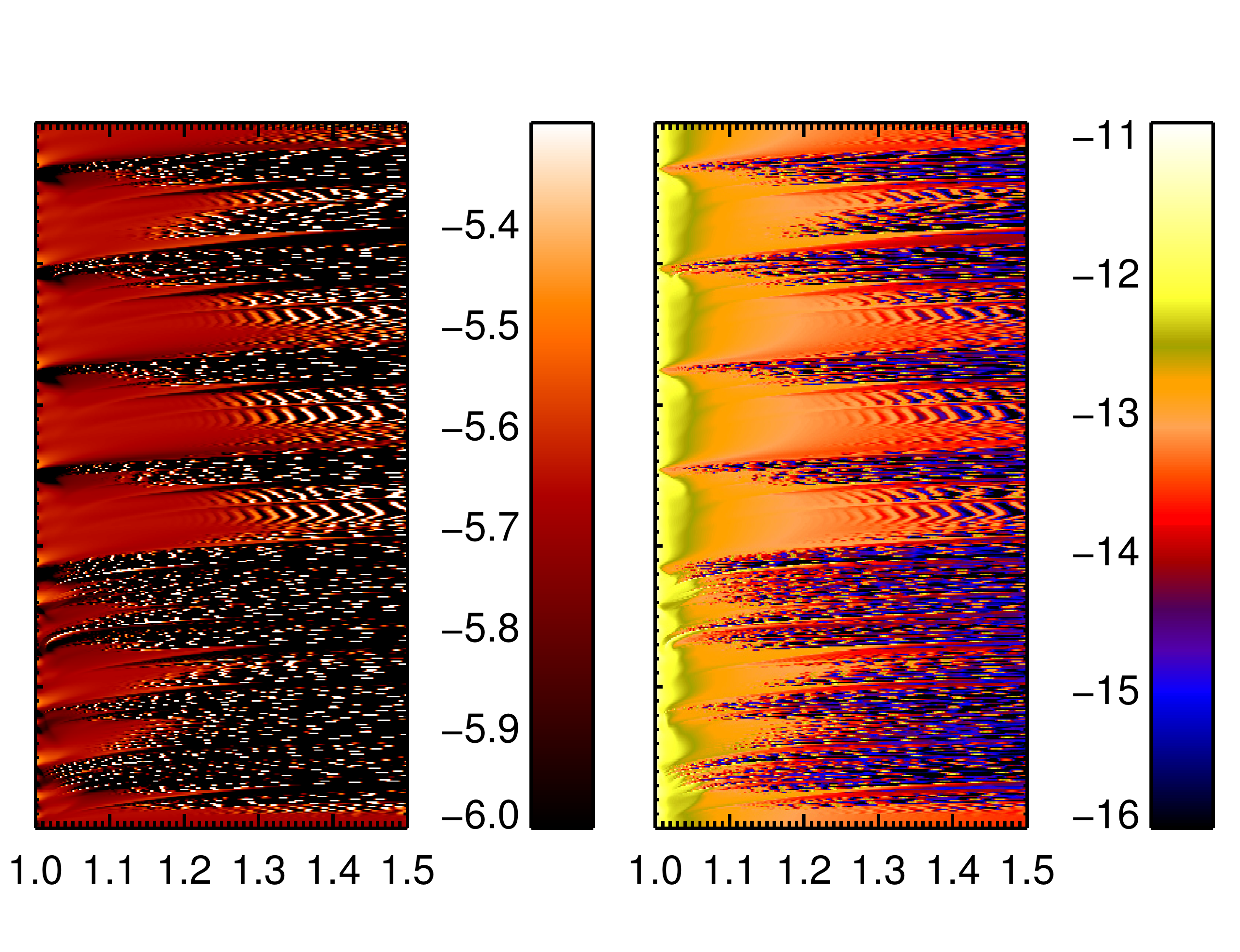}
            \centering
   \end{minipage} 
           
  \caption{Time-dependent SSF simulations computed with scattering source functions $S/S_{\rm th} =$ 1 (upper row), 0.95 (middle row), and 
  0.9 (lower row). All ordinates display time in ksec. From left to right the panels in each row show the ratio $a_{-}/a_{+}$, the topology function at the wind sonic point $f_{\rm top}(r_{\rm c})$, and color plots of mass-loss rate 
 ($\log dM/dt = \log \dot{M}$ [$M_\odot/\rm year$]) and density ($\log \rho$ [$g/cm^3$]), 
 where the abscissae show the radial coordinate in units of the stellar radius, $r/R_\star$. See text.}   
  \label{Fig:ssf}
\end{figure*}

\subsubsection{Fixed $S/S_{\rm thin}$ factors} 
\label{sec:sthin}

Figure~\ref{Fig:ssf} shows results from SSF simulations, using three simple
forms $S/S_{\rm thin} =$ 1, 0.95, and 0.9 for the source function 
that sets the scattering factor $s(r)$. (See equation~\ref{eq:sdef}.)
The first of these three cases (shown in the upper row of the figure) thus 
assumes an optically thin scattering source function from a uniformly 
bright stellar disk, while the second (middle row) 
and third (lower row) cases simply reduce this source function by respectively 5\% and 10\%. 
As we now demonstrate, such simple variation of the scattering term $s$ allows us to 
examine carefully the transition between nodal and  X-type topology in our numerical simulations. 

In the associated simulation results shown in Figure~\ref{Fig:ssf}, the left-most panels display the ratio
$a_{-}/a_{+}$ (equation~\ref{eq:apm}), computed every 1000~s at the mesh-point closest above the 
critical sonic point $r_{\rm c}$. For a given scattering term $s(r_{\rm c})$, the next panels 
plot the topology function $f_{\rm top}$ at this mesh-point. The remaining 
two panels display radius vs. time color-plots of mass-loss rate $\dot{M} \equiv 4 \pi r^2 \rho v$ 
and density. 

As evident from the figure, after an initial adjustment to the new 
force conditions, the standard simulation with $S/S_{\rm thin} =1$ relaxes to a stable wind 
base, wherein the line-drag of the diffuse radiation
\citep{Lucy84, Owocki85} exactly cancels the LDI at the stellar surface; in this case, 
self-excited structure develops only near the outer boundary at $r \approx 1.5R$ 
here. (See SO13 for structured simulations extending to larger radii.)

Building on the analysis in \S\ref{sec:sseq}, we can now interpret the relative stability of the base 
in terms of the $f_{\rm top}$ function at the sonic point. 
Even though $s(r_{\rm c}) < 0.5$ (reflecting the fact that the sonic point is located 
slightly above the stellar surface), the $a_{-}/a_{+}$ ratio yields 
$f_{\rm top} < 0$ at all times for this standard case (see first and second panels in top row), 
allowing then the wind around the sonic point to relax to a unique X-type solution.  

But in the runs with $S/S_{\rm thin} =$ 0.95 and 0.9, the lower values of $s(r_{\rm c})$ never 
allow the base to completely stabilize. At certain time-s-eps-converted-to.pdf $f_{\rm top}$ again becomes 
positive, implying a nodal wind topology. In the time-dependent simulations here, this gives rise to 
the same type of intrinsic global variability as found in the POC pure-absorption models,
and in the limb-darkened SSF simulations by SO13. 

Note that the run with $S/S_{\rm thin} = 0.95$ exhibits quite long periods where 
the wind appears almost stable, but as $f_{\rm top}$ approaches and 
exceeds unity, the wind reacts by admitting the nodal topology and its  
non-unique shallow-slope solutions. 
As compared to the transitional simulation with $S/S_{\rm thin} = 0.95$, 
the run with $S/S_{\rm thin} = 0.9$ has an even lower scattering term. 
After some initial adjustment-time, this now results in a strong and almost regular variability 
reaching all the way down to the wind base. Again times with $f_{\rm top} \ge 1$ 
correspond remarkably well to periods of strong variability, in between which 
the wind initially appears to relax, but then fails to settle down to a steady-state.

\subsubsection{Dependence on ion thermal speed}  
\label{sec:vthba}

As mentioned, the standard SSF models above all assume $v_{\rm th}/a = 0.28$. But the 
heuristic analysis in \S\ref{sec:tausobpm} predicts that as the Sobolev limit $v_{\rm th} \rightarrow 0$ is 
approached, obtaining a stable X-type solution requires a sonic point critical scattering 
term $s_{\rm c} > s_{\rm x} = 1/2$. And since an optically thin source function has 
$s_{\rm x} <1/2$ at the sonic point, this implies that as the ratio $v_{\rm th}/a$ is lowered,
even SSF models with $S/S_{\rm thin} = 1$ should eventually become nodal and 
so exhibit an unstable base. 

\begin{figure}
\begin{center}
\resizebox{\hsize}{!}  {\includegraphics[angle=90]{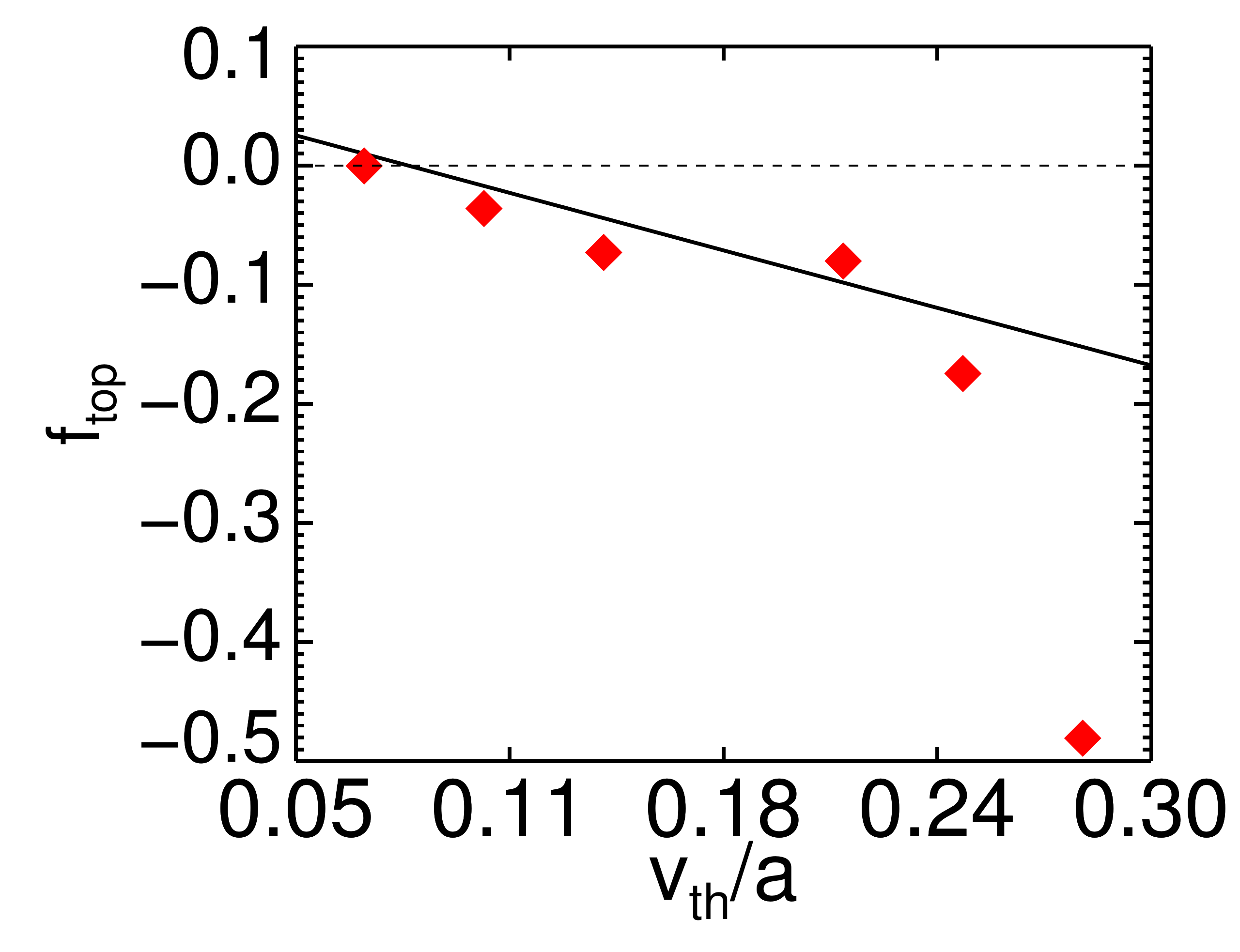}}
\caption{Sonic point topology function for different fixed ratios of ion thermal to sound speed. The red solid diamonds show results from SSF simulations with $S/S_{\rm thin} =1$ and the black solid line compares this to the heuristic analytic prediction eqn.~\ref{eq:ampapprox} using $s(r_{\rm c}) = 0.47$ (appropriate for simulations with $S/S_{\rm thin} =1$, see text). The dashed line marks the $f_{\rm top} = 0$ line where the wind transitions from X-type to nodal topology.}
\label{fig:tvth}
\end{center}
\end{figure}

\begin{figure}
\begin{center}
\resizebox{\hsize}{!}  {\includegraphics[angle=90]{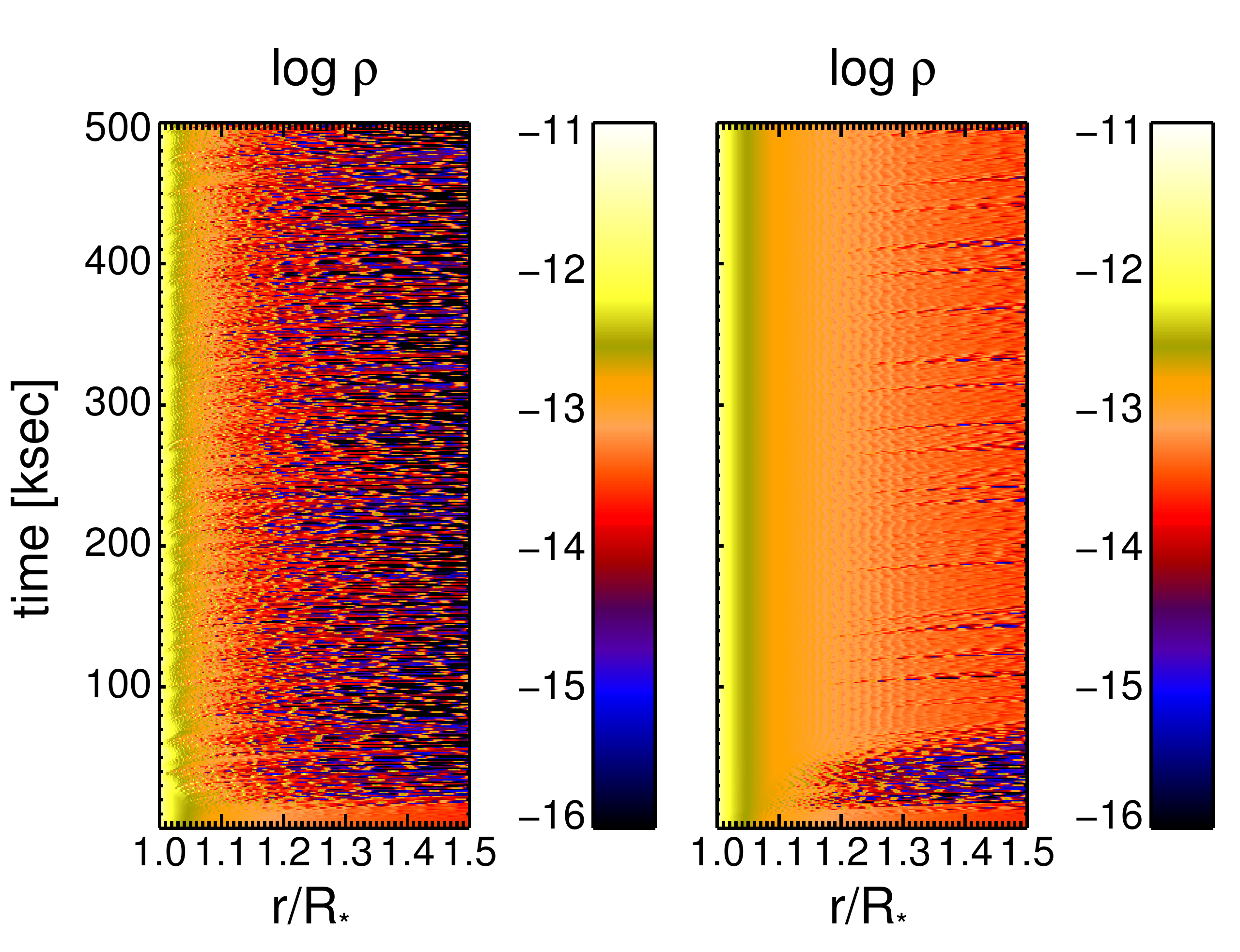}}
\caption{Density color-maps of two SSF simulations with $S/S_{\rm thin}=1$ and $v_{\rm th}/a$ = 0.05 (left) and 0.07 (right). See text.} 
\label{fig:rho_vth}
\end{center}
\end{figure}

To investigate this, we ran a set of standard SSF simulations with $S/S_{\rm thin} = 1$ but now with different $v_{\rm th}/a$ ratios. Figure \ref{fig:tvth} compares the asymptotic, steady-state values of $f_{\rm top}$ for these models to the simple analytic scaling equation (\ref{eq:ampapprox}) (using $s(r_{\rm c}) = 0.468$, appropriate for our standard optically thin source function). The figure shows that while the non-linear SSF simulations have larger escape asymmetries (i.e. more negative values of $f_{\rm top}$) at $v_{\rm th}/a = 0.28$, the numerical models agree quite well with the simple analytic scaling at lower thermal speeds. As anticipated, the simulations approach unstable conditions when the thermal speed is lowered, crossing the $f_{\rm top} = 0$ point that marks the transition from X-type to nodal topology at $v_{\rm th}/a \approx 0.07$. Inspections of simulations with $v_{\rm th}/a < 0.07$ further show that such models indeed exhibit a very chaotic wind base with extensive structure reaching all the way down to the lower boundary, as demonstrated by Fig.~\ref{fig:rho_vth}.  This figure plots density colour-maps of two models with $v_{\rm th}/a = 0.07$ and 0.05, illustrating how the first of these simulations (which has $f_{\rm top}$ very close to zero, see Fig.~\ref{fig:tvth}) is marginally stable, whereas the latter displays extensive structure already near the wind base. 

\subsubsection{Limb-darkened models}
\label{sec:limbdark} 

The SSF simulations above clearly demonstrate how scattering in the transonic wind region controls the admitted topology through the critical wind sonic point region. Building on these results, let us thus now 
examine the somewhat more physically realistic case of an Eddington-limb-darkened source function (like that adopted by SO13), returning again to the standard thermal speed with $v_{\rm th}/a=0.28$.

\begin{figure*}
    \vspace{1cm}
  \begin{minipage}{8.0cm}
      \includegraphics[angle=90,width=8.0cm]{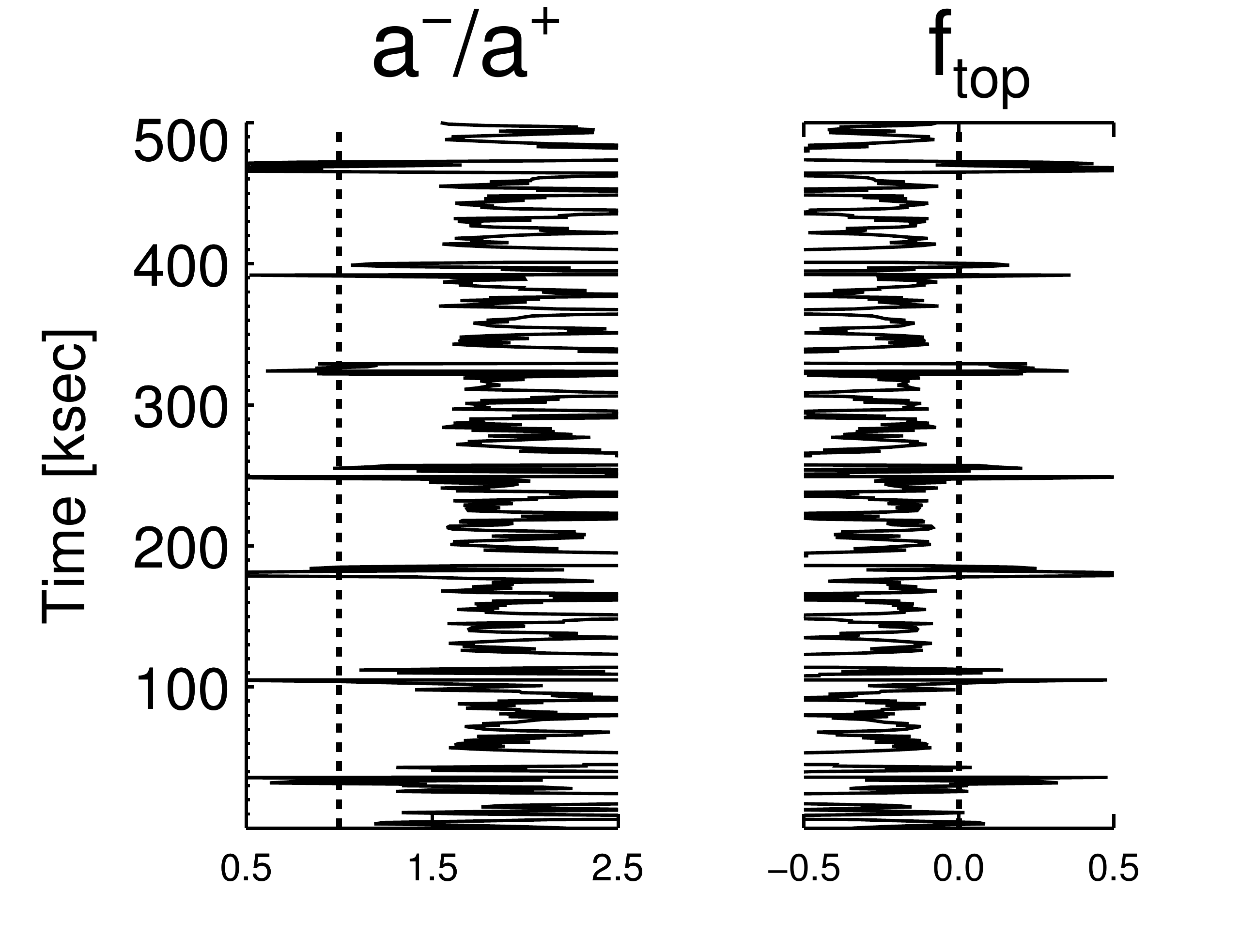}
    \centering
   \end{minipage}
    \begin{minipage}{8.0cm}
         \includegraphics[angle=90,width=8.0cm]{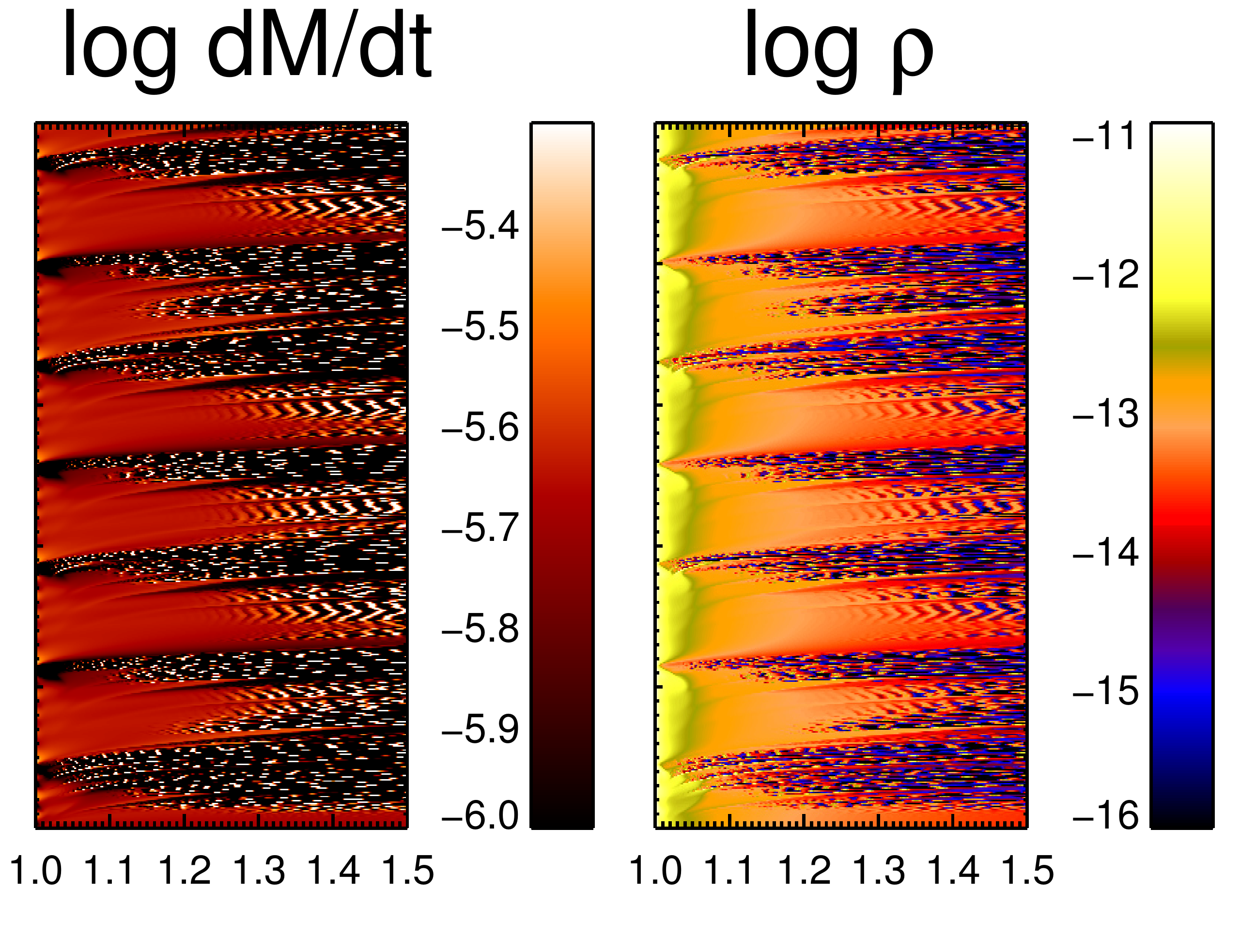}
            \centering
   \end{minipage} 
  \caption{Like Figure~\ref{Fig:ssf}, but for a limb-darkened source function.}   
  \label{Fig:ssf-ld}
\end{figure*}

Figure~\ref{Fig:ssf-ld} shows the same set of plots as in Figure~\ref{Fig:ssf}, but now for such a model. As discussed in \S 2, a limb-darkened source function naturally lowers the scattering term $s(r_{\rm c})$ and can thus easily reach $f_{\rm top} \ge 1$ and so nodal topology. And, indeed, Figure~\ref{Fig:ssf-ld} makes it immedietely clear that the strong variability extending down to the wind base in such limb-darkened SSF simulations is a direct consequence of $f_{\rm top}$ reaching values above unity during semi-regular intervals. As in the simulations above, this then gives rise to 
nodal topology solutions lying on the degenerate shallow-slope branch. Moreover, as 
discussed in detail in SO13, in such limb-darkened SSF simulations the lower scattering term 
also means the diffuse line-drag no longer cancels the LDI at the wind base, leading to non-linear 
feedback and presumably further enhancement of structure in these near-photospheric layers.  
 
\section{Discussion, Conclusions, and Outlook}
\label{sec:discussion}

The analysis here highlights the critical role played by the diffuse, scattered 
radiation field in the dynamics and variability of line-driven winds.

In a non-Sobolev model, in which the line force depends explicitly on velocity 
and radial coordinate (rather than on the velocity gradient), the sonic point 
constitutes a ``critical point" of vital importance for the wind dynamics. 
By extending here the pure-absorption analysis of POC, 
we show that standard, stable X-type wind solutions are admitted only in cases where 
the combination of the scattering term and the fore-aft asymmetry in 
escape is large enough. Within the SSF formalism, this is 
controlled by the sign of a sonic-point ``topology function" $f_{\rm top}$;
a negative  $f_{\rm top}$ gives X-type topology, whereas a positive $f_{\rm top}$ 
leads instead to nodal topology, with quite different behavior and 
an allowed branch of non-unique wind solutions. 

For the standard case of $v_{\rm th}/a = 0.28$ and an idealized optically thin scattering 
source function from a uniformly bright stellar disk, SSF simulations give 
numerically $f_{\rm top} < 0$, with then an associated relaxation to a 
stable wind base. But in cases where the source function is only 
mildly reduced from this idealized case (as in simple limb-darkened models), $f_{\rm top} $ 
becomes positive at certain times, implying a degenerate nodal topology. In the spherically 
symmetric simulations here, this leads to intrinsic structure and a semi-regular 
variability extending down to the wind base. This may be crucial for setting the
near-photospheric scale of various phenomena  observed in hot star winds 
(e.g., wind clumping and discrete-absorption components;  
\citealt{Eversberg98}; \citealt{Kaper99}; \citealt{Puls06}), and follow-up studies will 
examine how this variability manifests and evolves in multi-dimensional simulations, where 
the spherical shell structure is broken up by thin-shell and other hydrodynamic 
instabilities \citep{Dessart03, Dessart05}.  
 
The analysis here further illustrates quite vividly how very sensitive the wind dynamics 
is to details regarding the scattered radiation field in the transonic region. As such, it seriously
questions the validity of using Sobolev theory to compute the line force in this region, thus challenging
the Sobolev-based wind models often used in broad applications like stellar evolution \citep[e.g.,][]{Vink00}.
This motivates development of a new generation of models, able to quantitatively compute global wind properties accounting properly also for the diffuse component of the line force.

Indeed, full co-moving frame radiative transfer calculations typically find a significantly reduced source function in the transonic region (\citealt{Krticka10, Bouret12}; Sundvist et al., in prep.), and initial experimentation including a simplified escape-integral source function method (EISF, \citealt{Owocki99}) in \textit{time-dependent, dynamical} simulations shows significantly lower mass-loss rates and steeper velocity fields than predicted by Sobolev-based models \citep{Owocki99}. In view of the analysis here, it is interesting to note that while such EISF models certainly seem to lie on the nodal solution branch, they appear to relax to the unique steep solution, rather than admitting the degenerate shallow-slope solutions found in the SSF simulations here. 

Also the time-independent Monte-Carlo scattering models of the transonic flow by \citet{Lucy10}, find significantly lower mass-loss rates for main-sequence O-stars than predicted by the Monte-Carlo Sobolev simulations by \citet{Vink00}. However, in addition to never driving the wind beyond a few times the sound speed, the assumed pure-velocity scaling of the line-force in the \citet{Lucy07a, Lucy07b, Lucy10} models effectively suppresses appearance of a nodal topology, leaving instead only a single X-type solution (see discussion in \S\ref{sec:sonictop}).

Further work, e.g. using co-moving frame transfer to compute the radiation force, is thus needed to clarify which solution is obtained in dynamical simulations of line-driven winds, and under what circumstances the base of such winds relax to a steady state.  

\section*{Acknowledgments}

This work was supported in part by SAO Chandra grant TM3-14001A 
and NASA  Astrophysics Theory Program grant NNX11AC40G, awarded to the University of Delaware. 
We thank (soon to be) Dr. Dylan Kee for many fruitful discussions. 

\bibliographystyle{mn2e}
\bibliography{POC+SSF}

\begin{thebibliography}{39}
\expandafter\ifx\csname natexlab\endcsname\relax\def\natexlab#1{#1}\fi

\bibitem[{{Bouret} {et~al}\mbox{.}(2012){Bouret}, {Hillier}, {Lanz}, \&
  {Fullerton}}]{Bouret12}
{Bouret} J.-C., {Hillier} D.~J., {Lanz} T., {Fullerton} A.~W., 2012, \aap, 544,
  A67

\bibitem[{{Castor}, {Abbott} \& {Klein}(1975){Castor}, {Abbott}, \&
  {Klein}}]{Castor75}
{Castor} J.~I., {Abbott} D.~C., {Klein} R.~I., 1975, \apj, 195, 157

\bibitem[{{Castor}(1974)}]{Castor74}
{Castor} J.~L., 1974, \mnras, 169, 279

\bibitem[{{Cohen} {et~al}\mbox{.}(2011){Cohen}, {Gagn{\'e}}, {Leutenegger},
  {MacArthur}, {Wollman}, {Sundqvist}, {Fullerton}, \& {Owocki}}]{Cohen11}
{Cohen} D.~H., {Gagn{\'e}} M., {Leutenegger} M.~A., {MacArthur} J.~P.,
  {Wollman} E.~E., {Sundqvist} J.~O., {Fullerton} A.~W., {Owocki} S.~P., 2011,
  \mnras, 415, 3354

\bibitem[{{Cohen} {et~al}\mbox{.}(2014){Cohen}, {Wollman}, {Leutenegger},
  {Sundqvist}, {Fullerton}, {Zsarg{\'o}}, \& {Owocki}}]{Cohen14}
{Cohen} D.~H., {Wollman} E.~E., {Leutenegger} M.~A., {Sundqvist} J.~O.,
  {Fullerton} A.~W., {Zsarg{\'o}} J., {Owocki} S.~P., 2014, \mnras, 439, 908

\bibitem[{{Dessart} \& {Owocki}(2003)}]{Dessart03}
{Dessart} L., {Owocki} S.~P., 2003, \aap, 406, L1

\bibitem[{{Dessart} \& {Owocki}(2005)}]{Dessart05}
{Dessart} L., {Owocki} S.~P., 2005, \aap, 437, 657

\bibitem[{{Eversberg}, {Lepine} \& {Moffat}(1998){Eversberg}, {Lepine}, \&
  {Moffat}}]{Eversberg98}
{Eversberg} T., {Lepine} S., {Moffat} A.~F.~J., 1998, \apj, 494, 799

\bibitem[{{Feldmeier}, {Puls} \& {Pauldrach}(1997){Feldmeier}, {Puls}, \&
  {Pauldrach}}]{Feldmeier97}
{Feldmeier} A., {Puls} J., {Pauldrach} A.~W.~A., 1997, \aap, 322, 878

\bibitem[{{Gayley}(1995)}]{Gayley95}
{Gayley} K.~G., 1995, \apj, 454, 410

\bibitem[{{Holzer}(1977)}]{Holzer77}
{Holzer} T.~E., 1977, \jgr, 82, 23

\bibitem[{{Kaper} {et~al}\mbox{.}(1999){Kaper}, {Henrichs}, {Nichols}, \&
  {Telting}}]{Kaper99}
{Kaper} L., {Henrichs} H.~F., {Nichols} J.~S., {Telting} J.~H., 1999, \aap,
  344, 231

\bibitem[{{Krti{\v c}ka} \& {Kub{\'a}t}(2010)}]{Krticka10}
{Krti{\v c}ka} J., {Kub{\'a}t} J., 2010, \aap, 519, A50

\bibitem[{{Lucy}(1984)}]{Lucy84}
{Lucy} L.~B., 1984, \apj, 284, 351

\bibitem[{{Lucy}(2007{\natexlab{a}})}]{Lucy07a}
{Lucy} L.~B., 2007{\natexlab{a}}, \aap, 468, 649

\bibitem[{{Lucy}(2007{\natexlab{b}})}]{Lucy07b}
{Lucy} L.~B., 2007{\natexlab{b}}, \aap, 474, 701

\bibitem[{{Lucy}(2010)}]{Lucy10}
{Lucy} L.~B., 2010, \aap, 524, A41

\bibitem[{{Najarro}, {Hanson} \& {Puls}(2011){Najarro}, {Hanson}, \&
  {Puls}}]{Najarro11}
{Najarro} F., {Hanson} M.~M., {Puls} J., 2011, \aap, 535, A32

\bibitem[{{Owocki}(1991)}]{Owocki91}
{Owocki} S.~P., 1991, in IAU Symposium, Vol. 143, Wolf-Rayet Stars and
  Interrelations with Other Massive Stars in Galaxies, {van der Hucht} K.~A.,
  {Hidayat} B., eds., p. 155

\bibitem[{{Owocki}, {Castor} \& {Rybicki}(1988){Owocki}, {Castor}, \&
  {Rybicki}}]{Owocki88}
{Owocki} S.~P., {Castor} J.~I., {Rybicki} G.~B., 1988, \apj, 335, 914

\bibitem[{{Owocki} \& {Puls}(1996)}]{Owocki96}
{Owocki} S.~P., {Puls} J., 1996, \apj, 462, 894

\bibitem[{{Owocki} \& {Puls}(1999)}]{Owocki99}
{Owocki} S.~P., {Puls} J., 1999, \apj, 510, 355

\bibitem[{{Owocki} \& {Rybicki}(1984)}]{Owocki84}
{Owocki} S.~P., {Rybicki} G.~B., 1984, \apj, 284, 337

\bibitem[{{Owocki} \& {Rybicki}(1985)}]{Owocki85}
{Owocki} S.~P., {Rybicki} G.~B., 1985, \apj, 299, 265

\bibitem[{{Pauldrach}, {Puls} \& {Kudritzki}(1986){Pauldrach}, {Puls}, \&
  {Kudritzki}}]{Pauldrach86}
{Pauldrach} A., {Puls} J., {Kudritzki} R.~P., 1986, \aap, 164, 86

\bibitem[{{Pauldrach} {et~al}\mbox{.}(1994){Pauldrach}, {Kudritzki}, {Puls},
  {Butler}, \& {Hunsinger}}]{Pauldrach94}
{Pauldrach} A.~W.~A., {Kudritzki} R.~P., {Puls} J., {Butler} K., {Hunsinger}
  J., 1994, \aap, 283, 525

\bibitem[{{Poe}, {Owocki} \& {Castor}(1990){Poe}, {Owocki}, \&
  {Castor}}]{Poe90}
{Poe} C.~H., {Owocki} S.~P., {Castor} J.~I., 1990, \apj, 358, 199

\bibitem[{{Puls} {et~al}\mbox{.}(2006){Puls}, {Markova}, {Scuderi},
  {Stanghellini}, {Taranova}, {Burnley}, \& {Howarth}}]{Puls06}
{Puls} J., {Markova} N., {Scuderi} S., {Stanghellini} C., {Taranova} O.~G.,
  {Burnley} A.~W., {Howarth} I.~D., 2006, \aap, 454, 625

\bibitem[{{Puls}, {Springmann} \& {Lennon}(2000){Puls}, {Springmann}, \&
  {Lennon}}]{Puls00}
{Puls} J., {Springmann} U., {Lennon} M., 2000, \aaps, 141, 23

\bibitem[{{Puls}, {Vink} \& {Najarro}(2008){Puls}, {Vink}, \&
  {Najarro}}]{Puls08}
{Puls} J., {Vink} J.~S., {Najarro} F., 2008, \aapr, 16, 209

\bibitem[{{Rauw} {et~al}\mbox{.}(2015){Rauw}, {Herve}, {Naze},
  {Gonzalez-Perez}, {Hempelmann}, {Mittag}, {Schmitt}, {Schroeder}, {Gosset},
  {Eenens}, \& {Uuh-Sonda}}]{Rauw15}
{Rauw} G. {et~al.}, 2015, ArXiv e-prints

\bibitem[{{Smith}(2014)}]{Smith14}
{Smith} N., 2014, \araa, 52, 487

\bibitem[{{Sobolev}(1960)}]{Sobolev60}
{Sobolev} V.~V., 1960, {Moving envelopes of stars}. Cambridge: Harvard
  University Press, 1960

\bibitem[{{Sundqvist} \& {Owocki}(2013)}]{Sundqvist13}
{Sundqvist} J.~O., {Owocki} S.~P., 2013, \mnras, 428, 1837

\bibitem[{{Sundqvist}, {Puls} \& {Feldmeier}(2010){Sundqvist}, {Puls}, \&
  {Feldmeier}}]{Sundqvist10}
{Sundqvist} J.~O., {Puls} J., {Feldmeier} A., 2010, \aap, 510, 11

\bibitem[{{Sundqvist} {et~al}\mbox{.}(2011){Sundqvist}, {Puls}, {Feldmeier}, \&
  {Owocki}}]{Sundqvist11}
{Sundqvist} J.~O., {Puls} J., {Feldmeier} A., {Owocki} S.~P., 2011, \aap, 528,
  64

\bibitem[{{Sundqvist}, {Puls} \& {Owocki}(2014){Sundqvist}, {Puls}, \&
  {Owocki}}]{Sundqvist14}
{Sundqvist} J.~O., {Puls} J., {Owocki} S.~P., 2014, \aap, 568, 59

\bibitem[{{{\v S}urlan} {et~al}\mbox{.}(2013){{\v S}urlan}, {Hamann}, {Aret},
  {Kub{\'a}t}, {Oskinova}, \& {Torres}}]{Surlan13}
{{\v S}urlan} B., {Hamann} W.-R., {Aret} A., {Kub{\'a}t} J., {Oskinova} L.~M.,
  {Torres} A.~F., 2013, \aap, 559, A130

\bibitem[{{Vink}, {de Koter} \& {Lamers}(2000){Vink}, {de Koter}, \&
  {Lamers}}]{Vink00}
{Vink} J.~S., {de Koter} A., {Lamers} H.~J.~G.~L.~M., 2000, \aap, 362, 295

\end{thebibliography}

\newpage

\appendix

\section{Second-order Sobolev analysis}
\label{sec:appa}

\begin{figure}
\begin{center}
\includegraphics[scale=0.325]{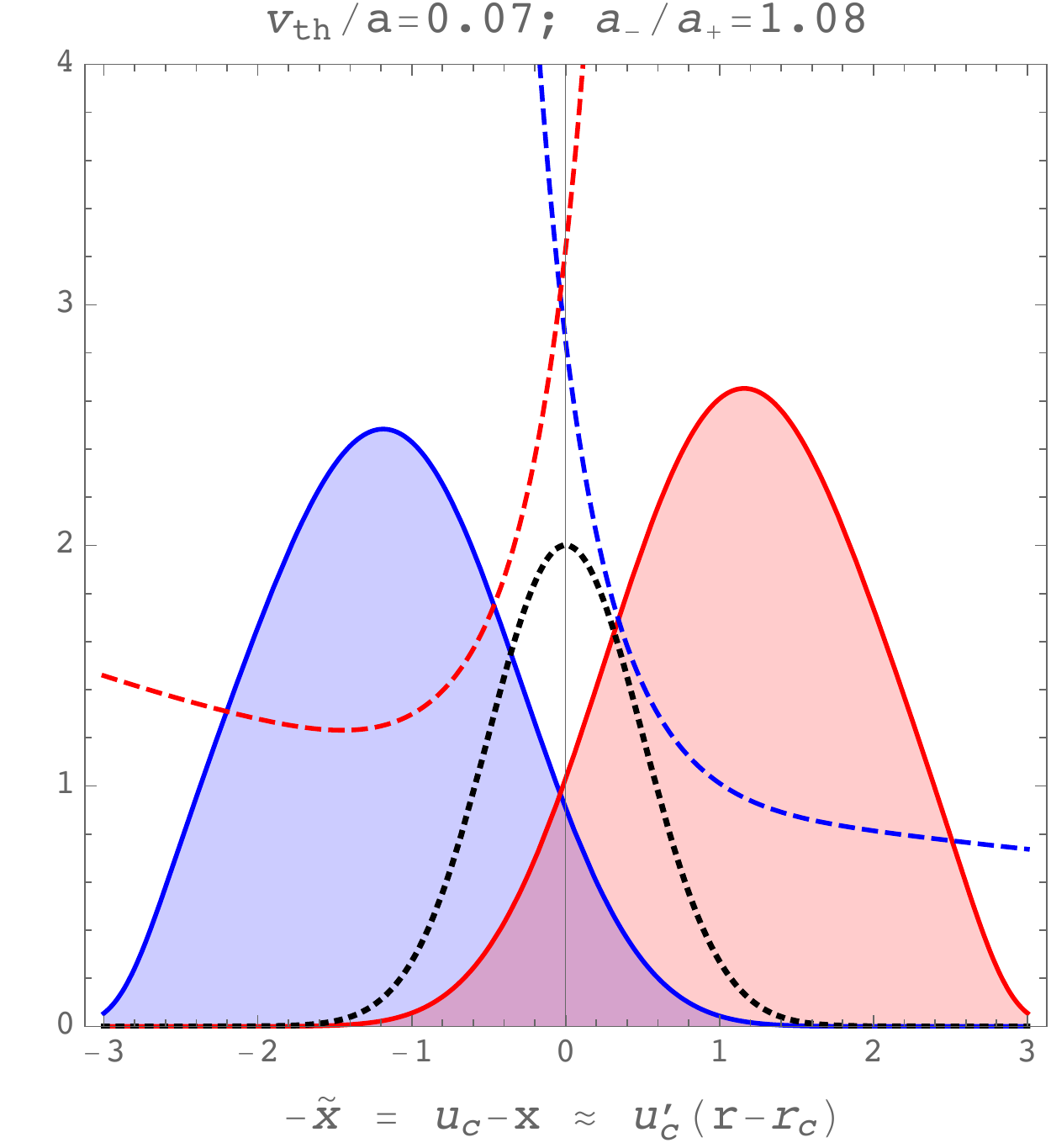}
\includegraphics[scale=0.325]{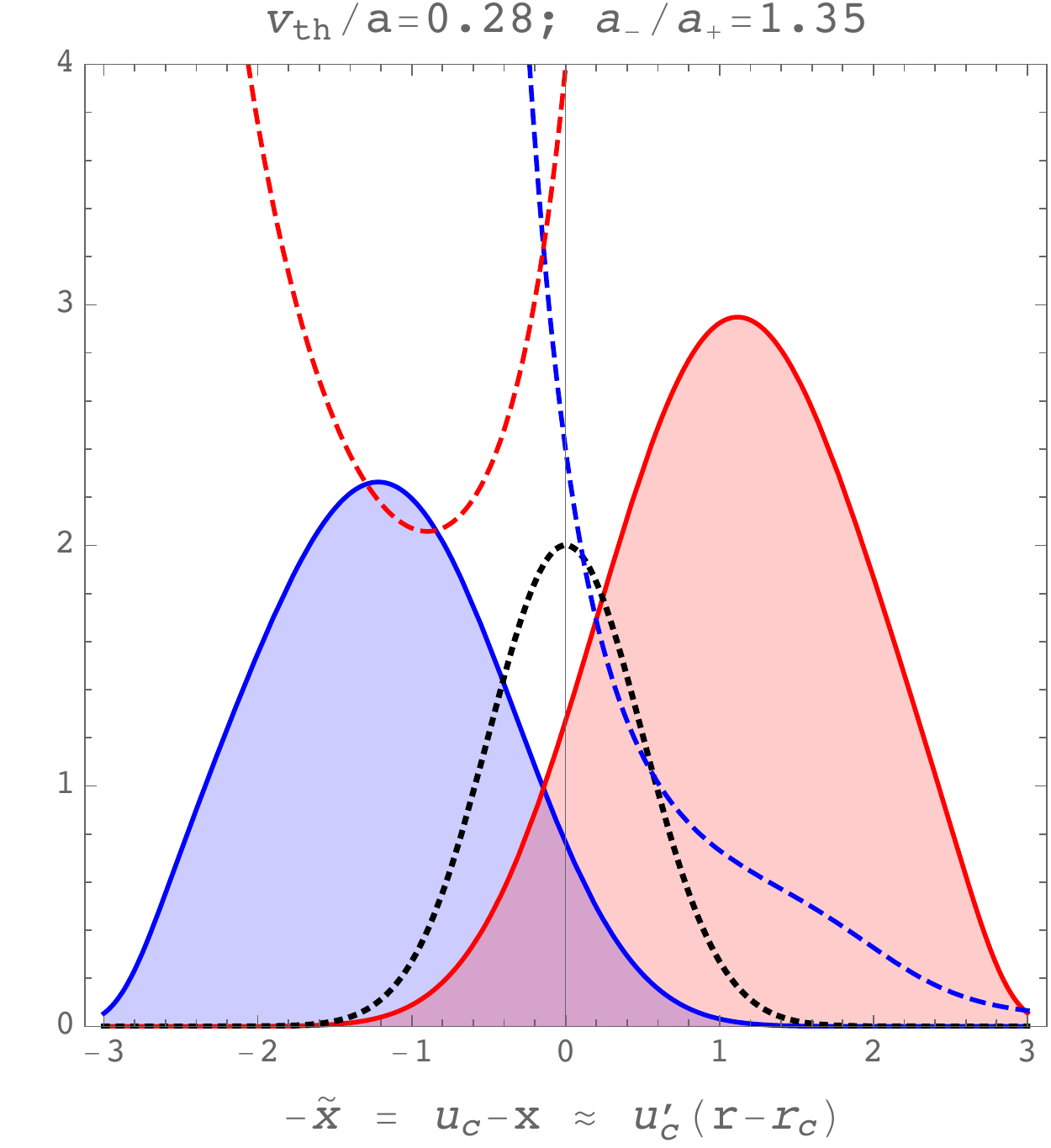}
\caption{
Terms from the integrand in the $a_\pm$ integral equation (\ref{eq:apm}), plotted vs. the negative of the critical point comoving frequency $- {\tilde x} = u_{\rm c} - x$,
for cases with 
$v_{\rm th}/a=0.07$  (left) and 
$v_{\rm th}/a=0.28$ (right).
The blue and red dashed curves show respectively forward/backward factors 
$(\tau_{\rm c}/t_{\rm c\pm})^{1+\alpha}$;
the dotted black curve shows $2\pi \phi^2(x-u_{\rm c})$; and the filled-area curves show the corresponding products of these, representing the full integrand of (\ref{eq:apm}).
Note that the red shaded area, representing $a_-$, is larger than the blue shaded area,  representing $a_+$, especially in the right panel with larger $v_{\rm th}/a$.
The headers in each panel label both the assumed value of $v_{\rm th}/a$  and  the computed value of $a_-/a_+$.
}
\label{fig:apm-int}
\end{center}
\end{figure}

To complement the simplified forward/backward optical depth estimates in \S\ref{sec:tausobpm}, let us now carry out a more formal ``second-order Sobolev'' analysis of the full optical depth integrals. 
For simplicity, we assume a purely radial ray with $y=0$ and thus $\mu_{\rm y} = 1$.
From equation (66) of OP96, we can write the forward-streaming optical depth at the critical point as
\beq
t_{\rm c+} (x) \equiv  t_{+} (x,0,r_{\rm c}) = 
\frac{\kappa_{\rm o}}{\kappa_{\rm max}} + 
\frac{\kappa_{\rm o}}{\kappa_{\rm e}}  \phi (x) + 
\Delta t_{\rm c+} (x)
\, ,
\eeq
where the first two terms account for the line-opacity cutoff (taken here to have a value $\kappa_{\rm max} \approx 10^6 \kappa_{\rm e}$) and the Shuster-Schwarzschild lower boundary condition at the stellar radius $R$;
the final term is given by the outward radial integral
\beqa
&&\Delta t_{\rm c+} (x)
\equiv
\int_{R}^{r_{\rm c}} \kappa_{\rm o} \rho(r) \phi (x -  u (r) ) \, dr
\\
~~~~~~~~&\approx& 
\kappa_{\rm o} \rho_{\rm c}  \int_{R}^{r_{\rm c}}
\left [ 1 + \frac{\rho_{\rm c}'}{\rho_{\rm c}} (r - r_{\rm c} ) \right  ] \phi (x -  u (r) )  dr
\\
&\approx& 
\kappa_{\rm o} \rho_{\rm c} 
\int_{x-u(R)}^{x- u_{\rm c}} 
\left [ 1 + \frac{v_{\rm c}'}{a} \, 
\frac{\tilde{x} - \tilde{x}_{\rm c} }{u_{\rm c}'} \right  ] \, \phi (\tilde{x} ) \, 
\frac{d \tilde{x}}{-u_{\rm c}'} 
\\
&\approx& 
\tau_{\rm c} 
\int_{x- u_{\rm c}}^{x}
\left [2 + \frac{v_{\rm th}}{a} \, 
\left ( \tilde{x} - x \right ) \right  ] \, \phi (\tilde{x} ) \, 
d \tilde{x}
\\
&\approx& 
\tau_{\rm c} 
\left [ 
\left (1 - x \frac{v_{\rm th}}{2a} \right ) 
 \left ( \erf(x) - \erf (x-u_{\rm c})  \right )
 \right . 
 \nonumber
\\
&&
~~~~~~~~~~~~~~~~~~~~~~~~~
+
\left .
 \frac{v_{\rm th}}{2a}
 \left (   \phi(x-u_{\rm c}) - \phi(x)  \right )
 \right ]
\, .
\label{eq:dtcp}
\eeqa
Here the intermediate forms assume a nearly linear variation in density and velocity within an approximately planar, steady-state flow near and below the critical (sonic) point.

The corresponding backward-streaming optical depth takes the form
 (cf.\ equation 68 of OP96), 
\beq
t_{\rm c-} (x) \equiv t_{-} (-x,0,r_{\rm c}) 
= \frac{\kappa_{\rm o}}{\kappa_{\rm max}} + \Delta t_{\rm c-} (x)
\, ,
\label{eq:tcm}
\eeq
where
\beqa
&&\Delta t_{\rm c-} (x)
\equiv
\int_{r_{\rm c}}^\infty \kappa_{\rm o} \rho(r) \phi (x -  u (r) ) \, dr
\\
~~ &\approx& 
\tau_{\rm c} 
\int_{-\infty}^{x- u_{\rm c}} 
\left [2 + \frac{v_{\rm th}}{a} \, 
\left ( \tilde{x} - x \right ) \right  ] \, \phi (\tilde{x} ) \, 
d \tilde{x}
\\
&\approx& 
\tau_{\rm c} 
\left [ 
\left (1 - x \frac{v_{\rm th}}{2a} \right ) 
\erfc (x-u_{\rm c})
 \right . 
-
\left .
 \frac{v_{\rm th}}{2a}
 \phi(x-u_{\rm c}) 
 \right ]
.
\label{eq:dtcm}
\eeqa

\begin{figure}
\begin{center}
\includegraphics[scale=0.325]{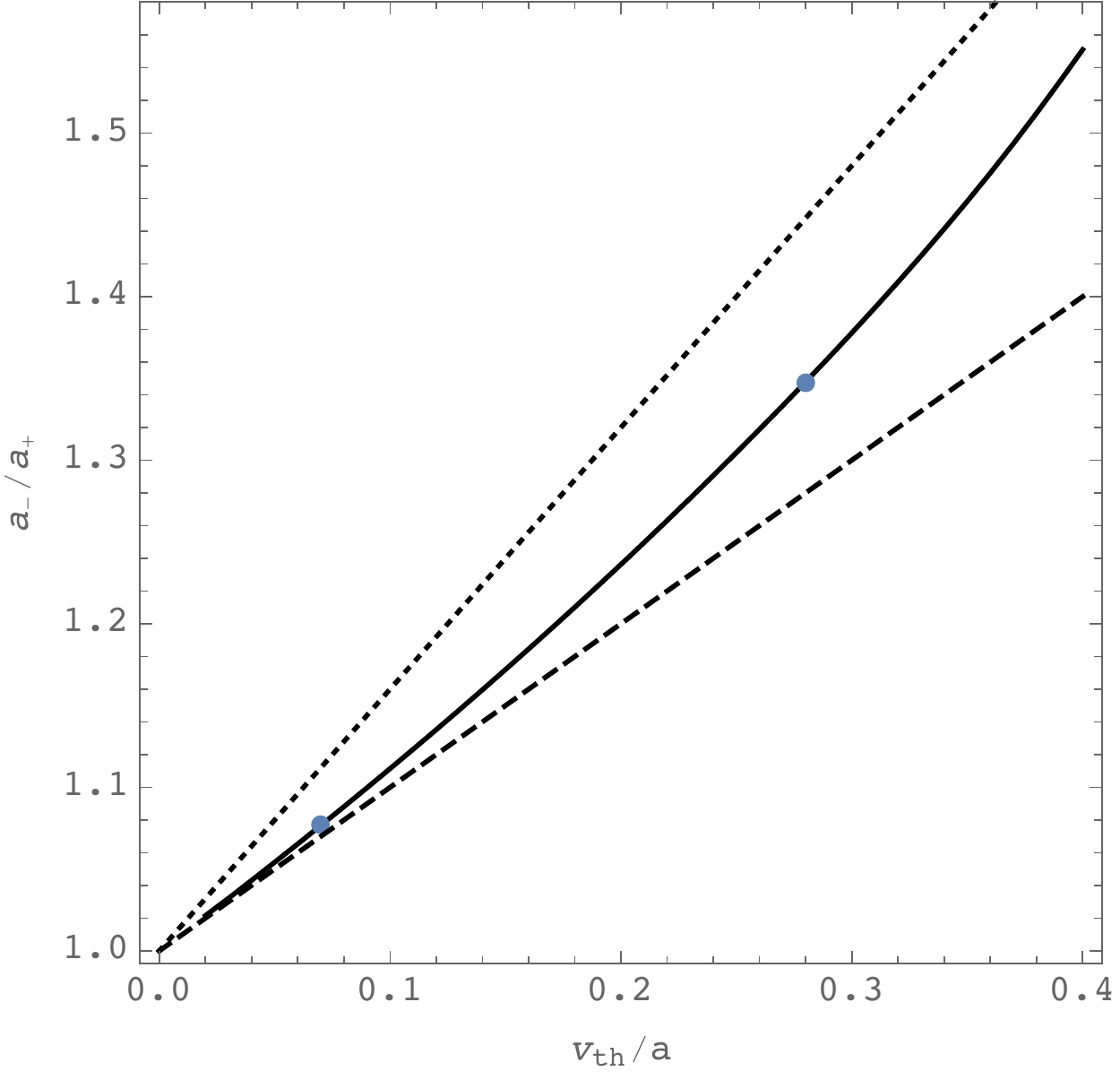}
\includegraphics[scale=0.325]{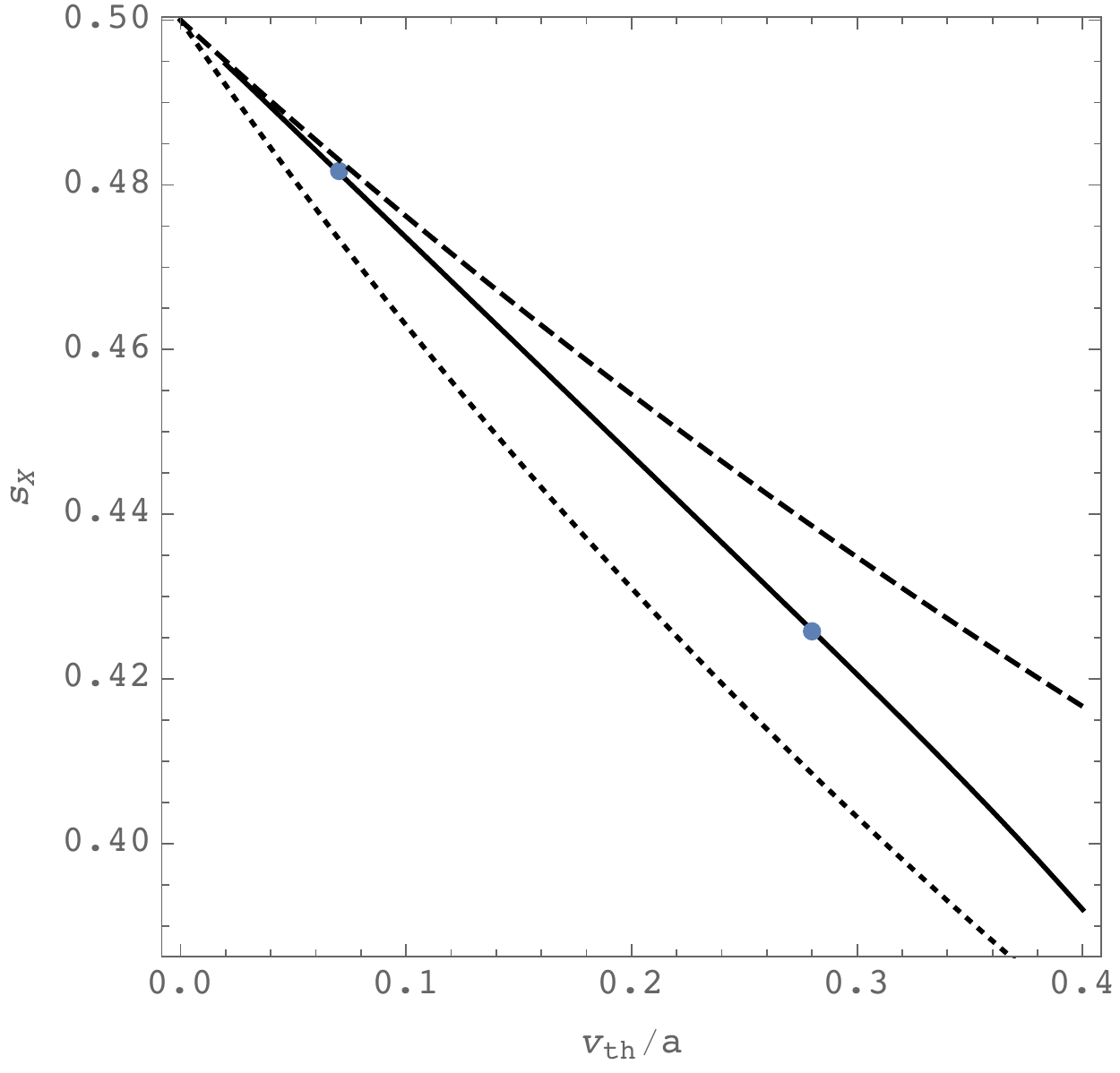}
\caption{
{\em Left:} 
The ratio $a_-/a_+$ plotted vs.\ $v_{\rm th}/a$ (solid curves), compared with the
analytic expansion form $a_-/a_+ \approx 1+ (1+\alpha) v_{\rm th}/a$ (dotted; cf.\ equation \ref{eq:ampapprox}) and the simple unit-slope linear form $a_-/a_+ \approx 1+ v_{\rm th}/a$ (dashed).
{\em Right:}
Associated variations of the minimum source function for an X-type topology, 
as defined by equation (\ref{eq:sxapm}).
In both panels, the filled circles highlight results for the left and right panels in figure \ref{fig:apm-int}.
}
\label{fig:apmvsvth}
\end{center}
\end{figure}

These analytic forms for $t_{\rm c\pm}$ can be used to compute the terms that determine the sign of $A$, which here depend on the full frequency integral forms of  equation (\ref{eq:apm}),
\beq
a_\pm \equiv
 \int_{-\infty}^{+\infty} \, \frac{\phi^2(x) }{t_{\rm c\pm}^{1+\alpha} (x)} \, dx
\, .
\label{eq:apm-int}
\eeq
To produce X-type critical solutions with $A>0$ , the critical-point scattering coefficient $s_{\rm c}$ must now exceed,
\beq
s_{\rm X} = \frac{1}{1+ \frac{a_-}{a_+}}
\, .
\label{eq:sxapm}
\eeq

Figure \ref{fig:apm-int} plots the frequency variation of the relevant integrand terms in the integral equation (\ref{eq:apm}) for $a_\pm$,  comparing results for $v_{\rm th}/a=0.07$ (left) vs.\ 
$v_{\rm th}/a=0.28$ (right).
Figure \ref{fig:apmvsvth} plots the 
ratio $a_-/a_+$ plotted vs.\ $v_{\rm th}/a$ (solid curves), with the filled circles highlighting results for the two cases shown in figure \ref{fig:apm-int}; the dotted and dashed curves compare simple 
linear forms with slopes of unity and $1+\alpha=1.65$, respectively.

For the standard case $v_{\rm th}/a = 0.28$ and $\alpha=0.65$, evaluation gives $a_-/a_+ \approx 1.35$ and $s_{\rm X} = 0.425$, results that are actually in quite good agreement with the more heuristic analysis in \S\ref{sec:tausobpm}.

\end{document}